\documentclass{article}

\usepackage{color}
\usepackage{graphicx}
\usepackage{amsthm}
\usepackage{amsmath}
\usepackage{amsfonts}
\usepackage{amssymb}
\usepackage{subcaption}
\usepackage{tikz}
\usepackage{hyperref}

\usepackage[normalem]{ulem}

\usepackage{geometry}
\newtheorem{theorem}{Theorem}[section]

\newtheorem{proposition}[theorem]{Proposition}

\newtheorem{definition}{Definition}

\def\l{\lambda}

\def\CC{\mathbb{C}}

\def\NN{\mathbb{N}}

\def\RR{\mathbb{R}}

\def\ZZ{\mathbb{Z}}

\def\cF{\boldsymbol{\mathcal{F}}}

\def\cP{\mathcal{P}}

\def\cS{\mathcal{S}}

\def\tl{\tilde{\l}}

\def\tG{\tilde{G}}
\def\BA{\boldsymbol{A}}
\def\BD{\boldsymbol{D}}
\def\BF{\boldsymbol{F}}
\def\BI{\boldsymbol{I}}

\def\BL{\boldsymbol{L}}
\def\BP{\boldsymbol{P}}
\def\BQ{\boldsymbol{Q}}
\def\BS{\boldsymbol{S}}
\def\BU{\boldsymbol{U}}
\def\BW{\boldsymbol{W}}

\def\Bf{\boldsymbol{f}}
\def\Bg{\boldsymbol{g}}

\def\Bu{\boldsymbol{u}}
\def\Bv{\boldsymbol{v}}
\def\Btu{\tilde{\boldsymbol{u}}}
\def\Bx{\boldsymbol{x}}
\def\By{\boldsymbol{y}}
\def\BvE{\boldsymbol{\mathcal{E}}}
\def\Bvphi{\boldsymbol{\phi}}
\def\Bvpsi{\boldsymbol{\psi}}
\def\Bphi{\hat{\boldsymbol{\phi}}}
\def\Bpsi{\hat{\boldsymbol{\psi}}}
\newcommand{\F}{\textup{\mbox{F}}}

\newcommand{\dist}{\textup{\mbox{dist}}}

\newcommand{\abs}[1]{\left| #1 \right|}

\newcommand{\norm}[1]{\left\lVert#1\right\rVert}

\title{Graph Convolutional Neural Networks via Scattering}
\author{Dongmian Zou, Gilad Lerman}

\begin{document}
\maketitle

\begin{abstract}
We  generalize the scattering transform to graphs and consequently construct a convolutional neural network on graphs. We show that under certain conditions, any feature generated by such a network is approximately invariant to permutations and stable to graph manipulations. Numerical results demonstrate competitive performance on relevant datasets.
\end{abstract}

\section{Introduction}\label{sec:intro}
Many interesting and modern datasets can be described by graphs. Examples include social \cite{LazPA09}, physical \cite{MasBB15}, and transportation \cite{ShuRV16} networks. The recent survey paper of Bronstein et al.~\cite{BroBL16} on geometric deep learning emphasizes the need to develop deep learning tools for such datasets and even more importantly to understand the mathematical properties of these tools, in particular, their invariances.

They also mention two types of problems that may be addressed by such tools. The first problem is signal analysis on graphs with applications such as classification, prediction and inference on graphs. The second problem is learning the graph structure with applications such as graph clustering and graph matching. 

Several recent works address the first problem \cite{BruZS13,EdwX16,DefBV16,HecCQ17}. In these works, the filters of the networks are designed to be parametric functions of graph operators, such as the graph adjacency and Laplacian, and the parameters of those functions have to be trained. 

The second problem is often explored with random graphs generated according to two common models: Erd{\H{o}}s--R{\'e}nyi, which is used for graph matching, and the Stochastic Block Model (SBM), which is used for community detection. Some recent graph neural networks have obtained state-of-the-art performance for graph matching \cite{NowVB17} and community detection \cite{BruL17,KipW16} with synthetic data generated from  the respective graph models. As above, the filters in these works are parametric functions of either the graph adjacency or Laplacian, where the parameters are trained. 

Despite the impressive progress in developing graph neural networks for solving these two problems, the performance of these methods is poorly understood. Of main interest is their invariance or stability to basic signal and graph manipulations. In the Euclidean case, the stability of a convolutional neural network \cite{GooBC16} to {rigid transformations and deformations} is best understood in view of the scattering transform \cite{Mal13}. The scattering transform has a multilayer structure and uses wavelet filters to propagate signals. It can be viewed as a convolutional neural network where no training is required to design the filters. Training is only required for the classifiers given the transformed data. Nevertheless, there is freedom in the selection and design of the wavelets. 
The scattering transform is approximately invariant to translation and rotation. More precisely, under strong assumptions on the wavelet and scaling functions and as the coarsest scale $-J$ approaches $-\infty$, the scattering transform becomes invariant to translations and rotations. Moreover, it is Lipschitz continuous with respect to smooth deformation. These properties are shown in \cite{Mal13} for signals in $L^2(\RR^d)$ and $L^2(H)$, where $H$ is a compact Lie group. 

It is interesting to note that the design of filters in existing graph neural networks is related to the design of wavelets on graphs in the signal processing literature. Indeed, the construction of wavelets on graphs use special operators on graphs such as the graph adjacency and Laplacian. As mentioned above, these operators are commonly used in graph neural networks.
The earliest works on graph wavelets \cite{CoiM06,MahM06} apply the normalized graph Laplacian to define the diffusion wavelets on graphs and use them to study multiresolution decomposition of graph signals. Hammond et al.~\cite{HamVG11} use the unnormalized graph Laplacian to define analogous graph wavelets and study properties of these wavelets such as reconstructibility and locality. One can easily construct a graph scattering transform by using any of these wavelets. A main question is whether this scattering transform enjoys the desired invariance and stability properties. 

In this work, we use a special instance of the graph wavelets of \cite{HamVG11} to form a graph scattering network and establish its covariance and approximate invariance to permutations and stability to graph manipulations. We also demonstrate the practical effectiveness of this transform in solving the two types of problems discussed above.

The rest of the paper is organized as follows. The scattering transform on graphs is defined in Section \ref{sec:wgcn}. Section \ref{sec:energy} shows that the full scattering transform preserves the energy of the input signal. This section also provides an absolute bound on the energy decay rate of components of the transform at each layer. Section \ref{sec:shift} proves the permutation covariance and approximate invariance of the {graph} scattering transform. It also briefly discusses {previously suggested candidates for the notion of translation or localization on graphs and} the possible covariance and approximate invariance of the scattering transform with respect to them. {Furthermore,
it clarifies why some special permutations are good substitutes for Euclidean rigid transformations.} Section \ref{sec:stability} establishes the stability of the scattering transform with respect to graph manipulations. Section \ref{sec:numerical} demonstrates competitive performance of the proposed graph neural network in solving the two types of problems. 

\section{Wavelet graph convolutional neural network}\label{sec:wgcn}
We first review the graph wavelets of \cite{HamVG11} in Section \ref{sec:wavelet}. We then use these wavelets and ideas of \cite{Mal13} to construct a graph scattering transform in Section \ref{sec:scattering}.  

\subsection{Wavelets on graphs}\label{sec:wavelet}
We review the wavelet construction of Hammond et al.~\cite{HamVG11} and adapt it to our setting. Our general theory applies to what we call simple graphs, that is, weighted, undirected and connected graphs with no self-loops. We remark that we may also address self-loops, but for simplicity we exclude them. Throughout the paper we fix an arbitrary simple graph $G = (V,E)$ with $N$ vertices. We also consistently use uppercase boldface letters to denote matrices and lowercase boldface letters to denote vectors or vector-valued functions. 

The weight matrix of $G$ is an $N \times N$ symmetric matrix $\BW$ with zero diagonal, where $\BW(n,m)$ denotes the weight assigned to the edge $\{n, m\}$ of G.
The degree matrix of $G$ is an $N \times N$ diagonal matrix with
\begin{equation}
\BD(n,n) = \sum_{m=1}^N \BW(n,m) ~, \quad 1 \leq n \leq N ~.
\end{equation}
The (unnormalized) Laplacian of $G$ is the $N \times N$ matrix 
\begin{equation}
\BL = \BD - \BW ~.
\end{equation}

The eigenvalues of $\BL$ are non-negative and the smallest one is $0$. Since the graph is connected, the eigenspace of $0$ (that is, the kernel of $\BL$) has dimension one. It is spanned by a vector with equal nonzero entries for all vertices. This vector represents a signal of the lowest possible ``frequency''. 

The graph Laplacian $\BL$ is symmetric and can be represented as
\begin{equation}\label{eq:eigendecomposition}
\BL = \sum_{l=0}^{N-1} \l_l \Bu_l \Bu_l^* ~,
\end{equation}
where $0 = \l_0 < \l_1 \leq \cdots \leq \l_{N-1}$ are the eigenvalues of $\BL$, $\Bu_0, \cdots, \Bu_{N-1}$ are the corresponding eigenvectors, and $*$ denotes the conjugate transpose. We remark that the phases of the eigenvectors of $\BL$ and their order within any eigenspace of dimension larger than 1 can be arbitrarily chosen without affecting our theory for the graph scattering transform formulated below.

Let $\Bf \in L_2(G)$ be a graph signal. Note that in our setting we can regard $L_2(G) \simeq L_2(V) \simeq \CC^N$, and without further specification we shall consider $\Bf \in \CC^N$. We define the Fourier transform $\cF: \CC^N \rightarrow \CC^N$ by
\begin{equation}\label{eq:fourierTransform}
\cF \Bf = \hat{\Bf} := \left( \Bu_l^* \Bf \right)_{l=0}^{N-1} ~,
\end{equation}
and the inverse Fourier transform $\cF^{-1}: \CC^N \rightarrow \CC^N$ by
\begin{equation}\label{eq:invFourierTransform}
\cF^{-1} \hat{\Bf} := \sum_{l=0}^{N-1} \hat{\Bf}(l) \Bu_l ~.
\end{equation}
Let  $\odot$ denote the Hadamard product, that is, for $\Bg_1$, $\Bg_2 \in \CC^N$, $\Bg_1 \odot \Bg_2 (l) = \Bg_1(l) \Bg_2(l)$, $l = 0, \cdots, N-1$. Define the convolution of $\Bf_1$ and $\Bf_2$ in $L_2(G)$ as the inverse Fourier transform of $\hat{\Bf_1} \odot \hat{\Bf_2}$, that is,
\begin{equation}\label{eq:defconvfg}
\Bf_1 \ast \Bf_2 = \cF^{-1} \left( \hat{\Bf}_1 \odot \hat{\Bf}_2 \right) = \sum_{l=0}^{N-1} \Bu_l  \hat{\Bf}_1 (l) \hat{\Bf}_2 (l) =\sum_{l=0}^{N-1} \Bu_l \Bu_l^* \Bf_1 \hat{\Bf_2}(l) =\sum_{l=0}^{N-1} \Bu_l \Bu_l^* \Bf_1 \Bu_l^* \Bf_2  ~.
\end{equation}
When emphasizing the dependence of $\ast$ on the graph $G$, we denote it by $\ast_G$. 

Euclidean wavelets use shift and scale in Euclidean space. For signals defined on graphs, which are discrete, the notions of translation and dilation need to be defined in the spectral domain. Hammond et al.~\cite{HamVG11} view $\RR$ as the spectral domain since it contains the eigenvalues of $\BL$. Their procedure assumes a scaling function $\phi$ and a wavelet functions $\psi$ \cite{Dau92,Mal99} with corresponding Fourier transforms $\hat{\phi}$ and $\hat{\psi}$. They have minimal assumptions on $\phi$ and $\psi$. In our construction, we consider dyadic wavelets, that is, 
\begin{equation}
\hat{\psi}_j (\omega) = \hat{\psi}(2^{-j} \omega), ~ j \in \ZZ~.
\end{equation} 
Also, we fix a scale $-J \in \ZZ$ of coarsest resolution and assume that $\phi$ and $\psi$ can be constructed from multiresolution analysis, that is,
\begin{equation}\label{eq:relationRhoPsi}
\abs{\hat{\phi}_{-J}}^2 + \sum_{j > -J} \abs{\hat{\psi}_j}^2 = 1 ~. 
\end{equation}
The graph wavelets of \cite{HamVG11} are constructed as follows in our setting. For $j > -J$, denote by $\Bpsi_j$ the vector in $\CC^N$ with the following entries: $\Bpsi_j(l) = \hat{\psi}_j(\l_l) = \hat{\psi}(2^{-j} \l_l)$, $l = 0, \cdots, N-1$.  Similarly, $\Bphi_{-J}(l) = \hat{\phi}_{-J}(\l_l) = \hat{\phi}(2^{-J} \l_l)$. In view of \eqref{eq:defconvfg},
\begin{equation}\label{eq:defconvpsi}
\Bf \ast \boldsymbol{\psi}_j = \sum_{l=0}^{N-1} \Bu_l \Bu_l^* \Bf \hat{\psi}(2^{-j} \l_l) \text{ for } j>-J \ \text{ and } \ \Bf \ast \boldsymbol{\phi}_{-J} = \sum_{l=0}^{N-1} \Bu_l \Bu_l^* \Bf \hat{\phi}(2^{J} \l_l)~.
\end{equation}
Note that $\Bf \ast \boldsymbol{\psi}_j$ and $\Bf \ast \boldsymbol{\phi}_{-J}$ are both in $\CC^N$. 
The graph wavelet coefficients of $\Bf \in L_2(G)$ are defined by 
\begin{equation}
\BQ_J \Bf := \Bf \ast \boldsymbol{\phi}_{-J} \ \text{ and } \BQ_j \Bf := \Bf \ast \boldsymbol{\psi}_j, ~ j > -J.
\end{equation} 
We use boldface notation for $\{ \BQ_j \}_{j \geq -J}$ to emphasize that they are operators even though the wavelet coefficients $\BQ_j \Bf (n), n = 1, \cdots, N$, are scalars. At last, we note that \eqref{eq:relationRhoPsi} implies that $\hat{\psi}(0)=0$. Combining this fact and \eqref{eq:defconvpsi} results in 
\begin{equation}
\label{eq:qj_orthogonal_to_1}
 \Bu_0^* \BQ_j = 0 \ \text{ for all } j > -J ~.
\end{equation}

\subsection{Scattering on graphs}\label{sec:scattering}
Our construction of convolutional neural networks on graphs is inspired by Mallat's scattering network \cite{Mal13}. As a feature extractor, the scattering transform defined in \cite{Mal13} is translation and rotation invariant when the coarsest scale approaches $-\infty$ and the wavelet and scaling functions satisfy some strong admissibility conditions. It is also Lipschitz continuous with respect to small deformations. The neural network representation of the scattering transform is the scattering network. It has been successfully used in image and audio classification problems \cite{BruM13}. 

We form a scattering network on graphs in a similar way, while using the graph wavelets defined above and the following definitions. A path $p = (j_1, \cdots, j_m)$ is an ordering of the scales of wavelets $j_1, \cdots, j_m > -J$. The length of the path $p$ is $\abs{p} = m$. The length of an empty path $p = \emptyset$ is zero. For a path $p = (j_1, \cdots, j_m)$ as above and a scale $j_{m+1} > -J$, we define the path $p +j_{m+1}$ as $p + j_{m+1} = (j_1, \cdots, j_m, j_{m+1})$. For a vector $\Bv \in \RR^n$ we denote $\abs{\Bv} = \left( \abs{\Bv(n)} \right)_{n=1}^N$ and note that the vectors $\Bv$ and $\abs{\Bv}$ have the same norm. 

For a scale $j > -J$, the one-step propagator $\BU[j]: \RR^N \rightarrow \RR^N$ is defined by
\begin{equation}\label{eq:defujf}
\BU[j]\Bf = \abs{\BQ_j \Bf} = \abs{\Bf \ast \boldsymbol{\psi}_j} = \left( \abs{\Bf \ast \boldsymbol{\psi}_j(n)} \right)_{n=1}^N , \quad \forall \Bf \in \RR^N ~.
\end{equation}

For $p \neq \emptyset$, the scattering propagator $\BU[p]: \RR^N \rightarrow \RR^N$ is defined by
\begin{equation}\label{eq:defUp}
\BU[p] = \BU[j_m] \BU[j_{m-1}] \cdots \BU[j_1] ~.
\end{equation}
For the empty path, we define $\BU[\emptyset] \Bf = \Bf$. We note that for any path $p$ and any scale $j_{m+1}>-J$ 
\begin{equation}\label{eq:pathappend}
\BU[p+j_{m+1}] = \BU[j_{m+1}] \BU[p] ~.
\end{equation}

The windowed scattering transform for a path $p$ is defined by
\begin{equation}\label{eq:defSf}
\BS[p]\Bf(n) = \BQ_J \BU[p]\Bf(n) = \BU[p]\Bf \ast \boldsymbol{\phi}_{-J} (n) = \sum_l \Bu_l \Bu_l^* \BU[p]\Bf(n) \hat{\boldsymbol{\phi}}(2^J \l_l) ~.
\end{equation}

Let $\Lambda^m$ denote the set of all paths of length $m \in \NN \cup \{0\}$, i.e., $\Lambda^m = \{p: \abs{p} = m \}$. The collection of all paths of finite length is denoted by $\cP_J := \bigcup_{m = 0}^{\infty} \Lambda^m$.
The scattering propagator and the scattering transform with respect to $\cP_J$, which we denote by $\BU[\cP_J]$ and $\BS[\cP_J]: \CC^N \rightarrow (\CC^N)^{\abs{\cP}}$ respectively, are defined as
\begin{equation}
\BU[\cP_J]\Bf = \left( \BU[p] \Bf \right)_{p \in \cP_J} ~\mbox{and }~ \BS[\cP_J]\Bf = \left( \BS[p] \Bf \right)_{p \in \cP_J} ~,~ \forall \Bf \in \CC^N ~.
\end{equation}
When we emphasize the dependence of the scattering propagator and transform on the graph $G$, we denote them by $\BU[G][\cP_J]$ and $\BS[G][\cP_J]$ respectively. A naturally defined norm on $\BU[\cP_J]$ and $\BS[\cP_J] \Bf$ is
\begin{equation}\label{eq:defUPSP}
\norm{\BU[\cP_J] \Bf} = \left( \sum_{p \in \cP_J} \norm{ \BU[p] \Bf }^2 \right)^{\frac{1}{2}} 
~ \mbox{and} ~
\norm{\BS[\cP_J] \Bf} = \left( \sum_{p \in \cP_J} \norm{ \BS[p] \Bf }^2 \right)^{\frac{1}{2}} ~,
\end{equation}
where $\norm{\cdot} = \norm{\cdot}_2$ denotes the $l_2$-norm on $\CC^N$. 

\begin{figure}[!ht]
\centering
\includegraphics[width=\linewidth]{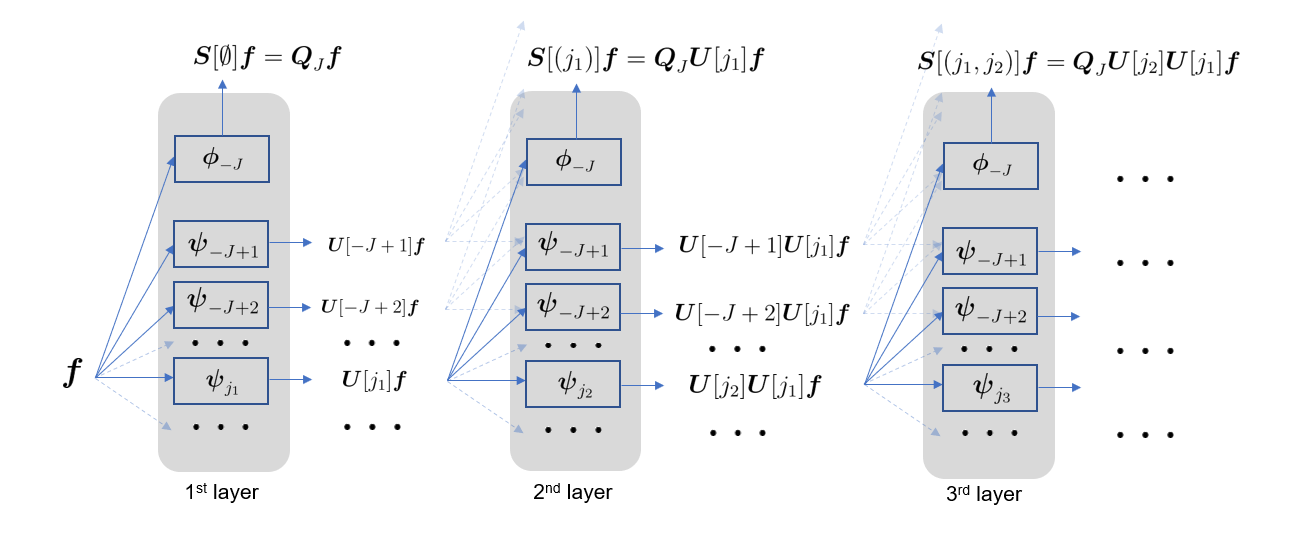}
\caption{Network representation of the scattering transform.}
\label{fig:cnn}
\end{figure}

In the terminology of deep learning, the scattering transform acts as a convolutional neural network on graphs. At the $m$-th layer, where $m \geq 0$,  the propagated signal is $\{\BU[p]\Bf: p \in \Lambda^m \}$ and the extracted feature is $\{\BS[p]\Bf: p \in \Lambda^m \}$. This network is illustrated in Figure \ref{fig:cnn}.

In a similar way, we can define the scattering transform for matrices of signals on graphs. Let $\BF = (\Bf_1, \Bf_2, \cdots, \Bf_D) \in \CC^{N \times D}$, where for each $1 \leq d \leq D$, $\Bf_d$  is a  complex signal of length $N$ on the same underlying graph. We define
\begin{equation}
\BS[\cP_J] \BF := \left( \BS[\cP_J] \Bf_d \right)_{d=1}^D
\end{equation}
and
\begin{equation}
\norm{\BS[\cP_J]\BF}_{\F} := \left( \sum_{d=1}^D \norm{\BS[\cP_J]\Bf_d}^2 \right)^{\frac{1}{2}} ~.
\end{equation}
Note that $\norm{\BS[p]\BF}_{\F}$ is the Frobenious norm of the matrix $\BS[p]\BF = \left( \BS[p]\Bf_d \right)_{d=1}^D$. Here and throughout the rest of the paper we denote by $\norm{\BA}_{\F}$ the Frobenius norm of a matrix $\BA$.

\section{Energy preservation}\label{sec:energy}

We discuss the preservation of energy of a given signal by the scattering transform. The signal is either $\Bf \in \CC^N$ with the energy $\norm{\Bf}^2$ or $\BF \in \CC^{N \times D}$ with the energy $\norm{\BF}_{\F}^2$. We first formulate our main result.

\begin{theorem}\label{thm:normPreserving}
The scattering transform is norm preserving. That is, for $\Bf \in \CC^N$ or $\BF \in \CC^{N \times D}$,
\begin{equation}
\label{eq:normPreserving}
\norm{\BS[\cP_J] \Bf}^2 = \norm{\Bf}^2 \mbox{ and } ~ \norm{\BS[\cP_J]\BF}_{\F}^2 = \norm{\BF}_{\F}^2 ~.
\end{equation}
\end{theorem}

The analog of Theorem \ref{thm:normPreserving} in the Euclidean case appears in \cite[Theorem 2.6]{Mal13}. However, the proof is different for the graph case. One basic observation analogous to the Euclidean case is the following.
\begin{proposition}\label{prop:energySplit}
For $\Bf \in \CC^N$ and $m \in \NN$,
\begin{equation}\label{eq:energyPreserve}
\sum_{p \in \Lambda^m} \norm{\BU[p] \Bf}^2 = \sum_{p \in \Lambda^{m+1}} \norm{\BU[p] \Bf}^2 + \sum_{p \in \Lambda^m} \norm{\BS[p] \Bf}^2 ~.
\end{equation}
\end{proposition}
This proposition can be rephrased as follows: the propagated energy at the $m$-th layer splits into the propagated energy at the next layer and the output energy at the current layer. In order to conclude Theorem \ref{thm:normPreserving} from Proposition \ref{prop:energySplit}, we quantify the decay rate of the propagated energy, which may be of independent interest. Fast decay rate means that few layers are sufficient to extract most of the energy of the signal. 
We define the decay rate of the scattering transform at a given layer as follows.
\begin{definition}
For $J \in \NN$, $m \in \NN$ and $r > 0$,  the  \emph{energy decay rate} of $\BS[\cP_J]$ at the $m$-th layer is $r$ if
\begin{equation}
\sum_{p \in \Lambda^{m+1}} \norm{\BU[p] \Bf}^2 \leq r \sum_{p \in \Lambda^m} \norm{\BU[p] \Bf}^2 ~.
\end{equation}
\end{definition}
In practice, different choices of graph $G$ and scale $J$ lead to different energy decay rates. Nevertheless, we establish the following generic result that applies to all graph scattering transforms under the construction in Section \ref{sec:scattering}.
\begin{proposition}\label{prop:layer}
The scattering transform $\BS[\cP_J]$ has energy decay rate of at least $1-2/N$ at all layers but the first one. This is the sharpest generic decay rate, though a better one can be obtained with additional assumptions on $J$, $\phi$, $\psi$ and $\BL$.
\end{proposition}
Note that in the Euclidean domain, no such generic result exists. Therefore, one has to choose the wavelets very carefully (see the admissibility condition in \cite[Theorem 2.6]{Mal13}). Numerical results illustrating the energy decay in the Euclidean domain are given in \cite{BruM13}. Furthermore,  theoretical rates are provided in \cite{WiaGB17} and \cite{CzaL17}, where \cite{WiaGB17} introduces  additional assumptions on the smoothness of input signals and the bandwidth of filters 
and \cite{CzaL17} studies time-frequency frames instead of wavelets. 

\begin{figure}[!ht]
    \centering
    \includegraphics[width=.5\linewidth]{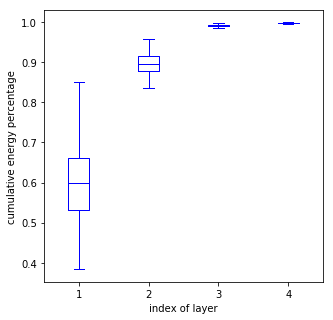}
 	\caption{Demonstration of fast energy decay rate for the graph scattering transform on MNIST. One hundred random images are drawn from the MNIST database, and the scattering transform is applied with the graph described in  Section \ref{sec:mnist}. The box plots summarize the distribution of the cumulative energy percentages for the random images.}
    \label{fig:energy}
\end{figure}

In practice, the propagated energy seems to decrease much faster than the generic rate stated in Proposition \ref{prop:layer}. Figure \ref{fig:energy} illustrates this claim. It considers $100$ randomly selected images from the MNIST database \cite{LecB98}. A graph that represents a grid of pixels shared by these images is used. Details of the graph and the dataset are described in Section \ref{sec:mnist}.
The figure reports the box plots of the cumulative percentage of the output energy of the scattering transform with $J=3$ for the first four layers and the 100 input images. That is, at layer $1 \leq m \leq 4$ the cumulative percentage for an image $\Bf$ is $\sum_{k=1}^m \sum_{p \in \Lambda^{k-1}} \norm{\BS[p] \Bf}^2 / \norm{f}^2$. 
We see that in the third layer, the scattering transform already extracts almost all the energy of the signal. Therefore, in practice we can estimate the graph scattering transform with a small number of layers, which is also evident in practice for the Euclidean scattering transform \cite{BruM13}.
 
\subsection{Proof of Proposition \ref{prop:energySplit}}
Application of (\ref{eq:defconvpsi}) and later (\ref{eq:relationRhoPsi}) implies that for any $\Bf \in \CC^N$
\begin{equation}\label{eq:fSingleSplit}
\begin{aligned}
~ & \sum_{j > -J} \norm{\Bf \ast \boldsymbol{\psi}_j}^2 + \norm{\Bf \ast \boldsymbol{\phi}_{-J}}^2 \\
~=~ & \sum_{j > -J} \norm{\sum_{l=0}^{N-1} \Bu_l \Bu_l^* \Bf \hat{\psi}(2^{-j} \l_l)}^2 + \norm{\sum_{l=0}^{N-1} \Bu_l \Bu_l^* \Bf \hat{\psi}(2^J \l_l)}^2 \\
~=~ & \sum_{j > -J} \sum_{l=0}^{N-1} \abs{\Bu_l^* \Bf \hat{\psi}(2^{-j}\l_l)}^2 + \sum_{l=0}^{N-1} \abs{\Bu_l^* \Bf \hat{\phi}(2^J \l_l)}^2 \\
~=~ & \sum_{l=0}^{N-1} \abs{\Bu_l^* \Bf}^2 \left( \sum_{j > -J} \abs{\hat{\psi}(2^{-j}\l_l)}^2 + \abs{\hat{\phi}(2^J \l_l)}^2 \right) \\
~=~ & \sum_{l=0}^{N-1} \abs{\Bu_l^* \Bf}^2 ~=~ \norm{f}^2 ~.
\end{aligned}
\end{equation}

Replacing $\Bf$ with $\BU[p]\Bf$, summing over all paths with length $m$ and applying (\ref{eq:pathappend}) yields
\begin{equation}\label{eq:pathbreakeq}
\begin{aligned}
\sum_{p \in \Lambda^m} \norm{\BU[p] \Bf}^2 ~=~ & \sum_{p \in \Lambda^m} \left( \sum_{j > -J} \norm{\BU[p] \Bf \ast \boldsymbol{\psi}_j}^2 +  \norm{\BU[p] \Bf \ast \boldsymbol{\phi}_J}^2 \right) \\
~=~ & \sum_{p \in \Lambda^m} \left( \sum_{j > -J} \norm{\abs{\BQ_j \BU[p] \Bf}}^2 +  \norm{\BS[p]\Bf}^2 \right) \\
~=~ & \sum_{p \in \Lambda^m} \sum_{j > -J} \norm{ \BU[p+j] \Bf}^2 + \sum_{p \in \Lambda^m}  \norm{\BS[p]\Bf}^2  \\
~=~ & \sum_{p \in \Lambda^{m+1}} \norm{\BU[p] \Bf}^2 + \sum_{p \in \Lambda^m} \norm{\BS[p] \Bf}^2 ~.
\end{aligned}
\end{equation}

\subsection{Proof of Proposition \ref{prop:layer}}
Recall that $\l_0 = 0$ and $\Bu_0 = \alpha/\sqrt{N} (1, \cdots, 1)^*$ where $\alpha \in \CC$ with $\abs{\alpha} = 1$. Note that (\ref{eq:relationRhoPsi}) implies that $\abs{\hat{\phi}(0)} = 1$. Note further that for any $p \in \Lambda^m$, $m \geq 1$, the entries of $\BU[p] \Bf$ are non-negative due to the absolute value in (\ref{eq:defujf}), and thus $\abs{\Bu_0^* \BU[p] \Bf} = \norm{\BU[p]\Bf}_1 / \sqrt{N}$. Consequently,
\begin{equation}
\label{eq:proof_best_rate}
\begin{aligned}
\norm{\BS[p]\Bf}^2 ~=~ & \norm{\BQ_J \BU[p]\Bf}^2 =  \norm{\sum_{l=0}^{N-1} \hat{\phi}(2^J \l_l) \Bu_l \Bu_l^* \BU[p]\Bf}^2 \\ 
~=~ & \sum_{l=0}^{N-1} \abs{ \hat{\phi}(2^J \l_l) \Bu_l^* \BU[p] \Bf}^2 
\geq  \abs{ \hat{\phi}(2^J \l_0) \Bu_0^* \BU[p] \Bf}^2 
=  \frac{1}{N} \norm{\BU[p] \Bf}_1^2~.
\end{aligned}
\end{equation}
Furthermore, we claim that 
\begin{equation}
\label{eq:simple_2_factor_ineq}
\norm{\BU[p] \Bf}_1^2 \geq  2 \norm{\BU[p] \Bf}^2 ~.
\end{equation}
Indeed, in view of \eqref{eq:qj_orthogonal_to_1}--\eqref{eq:defUp} and the form of $\Bu_0$, $\BU[p] \Bf = |\Bg|$, where $\Bg \in \CC^N$ satisfies $(1, \cdots, 1)^* \Bg=0$. One can easily show that the minimal value of $\norm{\Bg}_1^2$, over all $\Bg \in \CC^N$ satisfying $\norm{\Bg}^2=1$ and $(1, \cdots, 1)^* \Bg=0$, equals 2 and this concludes \eqref{eq:simple_2_factor_ineq}.

Combining \eqref{eq:proof_best_rate} and \eqref{eq:simple_2_factor_ineq} and summing the resulting inequality over $p \in \Lambda^m$ yields
\begin{equation}\label{eq:outMoreThanOneOverN}
\sum_{p \in \Lambda^m} \norm{\BS[p]\Bf}^2 \geq \frac{2}{N} \sum_{p \in \Lambda^m} \norm{\BU[p]\Bf}^2 ~.
\end{equation}
The combination of (\ref{eq:energyPreserve}) and (\ref{eq:outMoreThanOneOverN}) concludes the proof as follows
\begin{equation}
\sum_{p \in \Lambda^{m+1}} \norm{\BU[p] \Bf}^2 \leq \left( 1 - \frac{2}{N} \right) \sum_{p \in \Lambda^m} \norm{\BU[p] \Bf}^2 ~. 
\end{equation}

An improvement of this decay rate is possible if and only if one may strengthen the single inequality in \eqref{eq:proof_best_rate} and the inequality in \eqref{eq:simple_2_factor_ineq}. We show that these inequalities can be equalities for special cases and thus the stated generic decay rate is sharp. We first note that equality occurs in the inequality of \eqref{eq:proof_best_rate}  if, for example, $\hat{\phi}$ is the indicator function of $[0,2^J \lambda_1)$. Equality occurs in the second inequality when $\BU[p]\Bf$ has exactly two non-zero elements, for example, when $N=2$. 
These two cases can be simultaneously satisfied. 
We comment that certain choices of $J$, $\phi$, $\psi$ and $\BL$ imply different inequalities with stronger decay rates.

\subsection{Proof of Theorem \ref{thm:normPreserving}}
We write (\ref{eq:energyPreserve}) as
\begin{equation}
\sum_{p \in \Lambda^m} \norm{\BS[p] \Bf}^2 = \sum_{p \in \Lambda^m} \norm{\BU[p] \Bf}^2 - \sum_{p \in \Lambda^{m+1}} \norm{\BU[p] \Bf}^2
\end{equation}
and sum over $m \geq 0$, while recalling that $\BU[\emptyset]\Bf := \Bf$, to obtain that
\begin{equation}
\begin{aligned}
\label{eq:assist1_normPreserving}
\norm{\BS[\cP_J] \Bf}^2 ~=~ & \sum_{m \geq 0} \sum_{p \in \Lambda^m} \norm{\BS[p] \Bf}^2 \\
~=~ & \sum_{m \geq 0} \left( \sum_{p \in \Lambda^m} \norm{\BU[p] \Bf}^2 - \sum_{p \in \Lambda^{m+1}} \norm{\BU[p] \Bf}^2 \right) \\
~=~ & \lim_{m \rightarrow \infty} \left( \sum_{p \in \Lambda^0} \norm{\BU[p] \Bf}^2 - \sum_{p \in \Lambda^{m+1}} \norm{\BU[p] \Bf}^2 \right) \\
~=~ & \norm{\Bf}^2 - \lim_{m \rightarrow \infty} \sum_{p \in \Lambda^{m+1}} \norm{\BU[p] \Bf}^2 ~.
\end{aligned}
\end{equation}
Combining Proposition \ref{prop:layer} and (\ref{eq:fSingleSplit}) yields 
\begin{equation}
\label{eq:assist2_normPreserving}
\sum_{p \in \Lambda^{m+1}} \norm{\BU[p] \Bf}^2 \leq \left( 1-\frac{2}{N} \right)^m \sum_{p \in \Lambda^1} \norm{\BU[p] \Bf}^2 \leq  \left( 1-\frac{2}{N} \right)^m \norm{\Bf}^2 \rightarrow 0, ~\mbox{as}~ m \rightarrow \infty ~.        
\end{equation}
The first equality in \eqref{eq:normPreserving} clearly follows from \eqref{eq:assist1_normPreserving}
and \eqref{eq:assist2_normPreserving}. The second equality in \eqref{eq:normPreserving} is an immediate consequence of the first equality and the observation that for $\BF = (\Bf_1, \cdots, \Bf_D)$, $\norm{\BF}_{\F}^2 = \sum_{d=1}^D \norm{\Bf_d}^2$.

\section{Permutation covariance and invariance}\label{sec:shift}

When applying a transformation $\Phi[G]$ to a graph signal it is natural to expect that relabeling the graph vertices and the corresponding signal's indices before applying the transformation has the same effect as relabeling the corresponding indices after applying the transformation. More precisely, let $\BP \in S_N$ be a permutation, where $S_N$ denotes the symmetric group on $N$ letters, then it is natural to ask whether  
\begin{equation}
\label{eq:cov_perm}
\Phi[\BP G](\BP \Bf) = \BP \Phi[G](\Bf). 
\end{equation}
In deep learning, the property expressed in \eqref{eq:cov_perm} is referred to as covariance to permutations. On the other hand, invariance to permutations means that  
\begin{equation}
\label{eq:inv_perm}
\Phi[\BP G] (\BP \Bf) = \Phi[G] (\Bf).
\end{equation}

Ideally, a graph-based classifier should not be sensitive to ``graph-consistent relabeling'' of the signal coordinates. The analog of this ideal request in the Euclidean setting is that a classifier of signals defined on $\RR^D$ should not be sensitive to their {rigid transformations}. In the case of classifying graph signals by first applying a feature-extracting transformation and then a standard classifier,  this ideal request translates to permutation invariance of the initial transformation. However, permutation invariance is a very strong property that often contradicts the necessary permutation covariance. We show here that the scattering transform is permutation covariant and if the scaling function is sufficiently smooth and $J$ approaches infinity, then it becomes permutation invariant. 

We first exemplify the basic notions of covariance and invariance in Section \ref{subsec:example}. Section~\ref{subsec:main_results_cov_inv} reviews the few existing results on permutation covariance and invariance of graph neural networks and then presents our results for the scattering network.  
{Section \ref{subsec:graphtranslation} explains why some permutations are natural generalizations of rigid transformations and then discusses previous broad generalizations of the notion of ``graph translation'' and their possible covariance and invariance properties.}
Sections \ref{subsec:proofpermcov} and \ref{subsec:prooftransinv} prove the main results formulated in Section~\ref{subsec:main_results_cov_inv}.

\subsection{Basic examples of graph permutations, covariance and invariance}
\label{subsec:example}
\begin{figure}[!ht]
\centering
\begin{subfigure}{.32\textwidth}
	\centering
    \includegraphics[width=\linewidth]{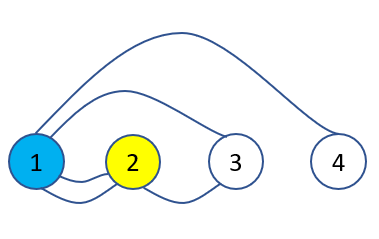}
    \caption{$(G, \Bf)$}
    \label{fig:permIll1}
\end{subfigure}
\begin{subfigure}{.32\textwidth}
	\centering
    \includegraphics[width=\linewidth]{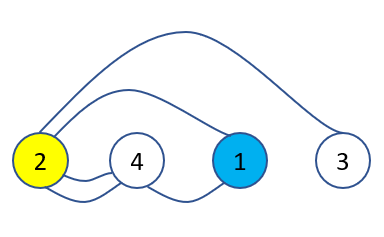}
    \caption{$(\BP G, \Bf)$}
    \label{fig:permIll2}
\end{subfigure}
\begin{subfigure}{.32\textwidth}
	\centering
    \includegraphics[width=\linewidth]{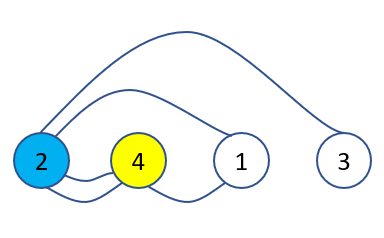}
    \caption{$(\BP G, \BP \Bf)$}
    \label{fig:permIll3}
\end{subfigure}
\caption{Illustration of permutation for a particular example of a graph and signal discussed in this section.}
\label{fig:permIll}
\end{figure}

For demonstration, we focus on the graph $G$ depicted in Figure \ref{fig:permIll1}. In this graph, each drawn edge has weight one and thus the double edge between the first two nodes has the total weight 2. The weight matrix of the graph is 
\begin{equation}\label{eq:defWeightMatrix}
\BW = \begin{bmatrix}
    0 & 2 & 1 & 1 \\
    2  & 0 & 1 & 0 \\
   1  & 1 & 0 & 0 \\
   1 & 0 & 0 & 0 
\end{bmatrix}
.
\end{equation}
The signal $\Bf = (2,1,0,0)^*$ is depicted on the graph with different colors corresponding to different values.
The following permutation is applied to the graph in Figure \ref{fig:permIll2}:
\begin{equation}\label{eq:defPermMatrix}
\BP = \begin{bmatrix}
0	&0	&1	&0 \\
1	&0	&0	&0 \\
0	&0	&0	&1 \\
0	&1	&0	&0 
\end{bmatrix}
.
\end{equation}
Figure \ref{fig:permIll3} applies the permutation both to the signal and the graph.

An example of a transformation $\Phi[G]$ can be the replacement of the signal values in the two vertices connected by the edge of weight 2. This transformation is independent of the labeling of the graph and is thus permutation covariant. This can be formally verified as follows, while using for simplicity the permutation $\BP$ defined in (\ref{eq:defPermMatrix}). For a signal $\Bf = (f_1, f_2, f_3, f_4)^*$, $\BP \Bf = (f_3, f_1, f_4, f_2)^*$. Furthermore, $\Phi[G]$ swaps the first two entries of a signal, while $\Phi[\BP G]$ swaps the second and the fourth entries (the second claim is obvious from Figure \ref{fig:permIll2}). Accordingly, $\Phi[G] (\Bf) = (f_2, f_1, f_3, f_4)^*$ and $\Phi[\BP G] (\BP \Bf) = (f_3, f_2, f_4, f_1)^*$. One can readily check that indeed $\BP \Phi[G] (\Bf) = (f_3, f_2, f_4, f_1)^* = \Phi[\BP G] (\BP \Bf)$.

Another example is the transformation $\Phi[G] (\Bf) = \BW[G] \Bf$, where $\BW[G] \equiv \BW$  is the weight matrix in (\ref{eq:defWeightMatrix}). This transformation is also independent of the labeling of the graph and thus permutation covariant. This property can also be formally verified as follows:
\begin{equation}
\Phi[\BP G] (\BP \Bf) = \BW[\BP G] \BP \Bf = \BP \BW[G] \BP^* \BP \Bf = \BP \BW[G] \Bf = \BP \Phi[G](\Bf) ~.
\end{equation}
Similarly, $\Phi[G] (\Bf) = \BL[G] \Bf$, where  $\BL[G]$ is the graph Laplacian, is permutation covariant.

The above three examples of permutation covariant transformations are not permutation invariant. An example of a permutation invariant transformation $\Phi[G]$, but not permutation covariant, maps the signal $\Bf = (f_1, \ldots, f_4)^*$ to the signal $\Phi[G] \Bf = (\max_{i=1}^4 f_i, 0,0,0)^*$. Clearly the output $\Phi[G] \Bf$ is not affected by permutation of the input signal and is thus permutation invariant. On the other hand, zeroing out three specified signal coordinates, instead of three vertices with unique graph properties (e.g., the vertices connected by at least two edges), violates permutation covariance. 

The latter example demonstrates in a very simplistic way the value of invariance for classification. Indeed, assume that there are two types of signals with low and high values and a classifier tries to distinguish between the two classes according to the first coordinate of $\Phi[G] (\Bf)$ by checking whether it is larger than a certain threshold or not. Then this procedure can distinguish the two types of signals without getting confused with signal relabeling. Permutation covariance does not play any role in this simplistic setting, since the classifier only considers the first coordinate of $\Phi[G] (\Bf)$ and ignores the rest of them. 

\subsection{Permutation covariance and invariance of graph neural networks}
\label{subsec:main_results_cov_inv}
The recent works of Gilmer et al.~\cite{GilSR17} and Kondor et al.~\cite{KonS18} discuss permutation covariance and invariance for composition schemes on graphs, where message passing is a special case. Composition schemes are covariant to permutations since they do not depend on any labeling of the graph vertices. Moreover, if the aggregation function {of the composition scheme} is invariant to permutations, so is the whole scheme \cite[Proposition 2]{KonS18}. {However, aggregation leads to loss of local information, which might weaken the performance of the scheme.} 

Methods based on graph operators, such as the graph adjacency, weight or Laplacian, are not invariant 
to permutations (see demonstration in Section \ref{subsec:example}). Nevertheless, the scattering transform is approximately permutation invariant when the wavelet scaling function is sufficiently smooth. Furthermore, when $J$ approaches infinity it becomes invariant to permutations. We first formulate its permutation covariance and then its approximate permutation invariance.

\begin{proposition}\label{prop:permCov}
Let $G$ be a simple graph and $\BS[G][\cP_J]$ be the graph scattering transform with respect to $G$. For any $\Bf \in \CC^N$ and $\BP \in \cS_N$, 
\begin{equation}\label{eq:permCov}
\BS[\BP G][\cP_J] \BP \Bf = \BP \BS[G][\cP_J] \Bf ~.
\end{equation}
\end{proposition}

\begin{theorem}\label{thm:transinv}
Let $G$ be a simple graph and $\BS[G][\cP_J]$ be the graph scattering transform with respect to $G$. Assume that the Fourier transform of the scaling function $\phi$ of $\BS[G][\cP_J]$ decays as follows: $\hat{\phi}(\omega)  \leq C_{\phi} /|\omega|$, where $C_{\phi}$ is a constant depending on $\phi$. For any $\Bf \in \CC^N$ and $\BP \in \cS_N$
\begin{equation}
\label{eq:approx_inv}
\norm{\BS[\BP G][\cP_J] \BP \Bf - \BS[G][\cP_J] \Bf} \leq C_{\phi} \, 2^{-(J+0.5)}  \, \l_1^{-1} \, \sqrt{N+2} \norm{\BP-\BI} \norm{\Bf} ~.
\end{equation}
In particular, the scattering transform is invariant as $J$ approaches infinity.
The result also holds if $\Bf \in \CC^N$ is replaced with $\BF \in \CC^{N \times D}$ and the Euclidean norm is replaced with the Frobenius norm.
\end{theorem}

\subsection{Generalized graph translations}\label{subsec:graphtranslation}

{Permutation invariance on graphs is an important notion, which is motivated by concrete applications \cite{GutVW16,DuvMI15}. It can be seen as an analog of translation invariance in Euclidean domains, which is also essential for applications \cite{Mal13, GooBC16}.} {A different line of research asks for the most natural notion of translation on a graph \cite{ShuRV16, SanM13}. We show here that very special permutations of signals on graphs naturally generalize the notion of translation or rigid transformation of a signal in a Euclidean domain. More precisely, there is a planar representation of the graph on which the permutation acts like a rigid transformation. However, in general, there are many permutations that act very differently than translations or rigid transformations in a Euclidean domain, though, they still preserve the graph topology. Indeed, the underlying geometry of general graphs is richer than that of the Euclidean domain.} 
We later discuss {previously suggested} generalized notions of ``translations'' on graphs and the possible covariance and invariance of a modified graph scattering transform with respect to these.  

\begin{figure}[!ht]
\centering
\begin{subfigure}{.25\textwidth}
	\centering
    \includegraphics[width=\linewidth]{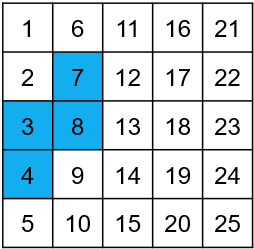}
    \caption{$(G, \Bf)$ \\ ~ }
    \label{fig:permIsTransl1}
\end{subfigure}
{\large$\xrightarrow{\textup{~relabel~}}$}%
\begin{subfigure}{.25\textwidth}
	\centering
    \includegraphics[width=\linewidth]{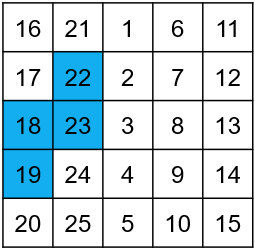}
    \caption{$(\BP G, \BP \Bf)$ \\ ~ }
    \label{fig:permIsTransl2}
\end{subfigure}
{\large$\xrightarrow{\textup{rearrange}}$}%
\begin{subfigure}{.25\textwidth}
	\centering
    \includegraphics[width=\linewidth]{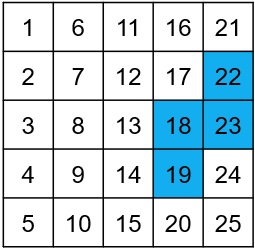}
    \caption{Indices of $(\BP G, \BP \Bf)$ \\ rearranged as in (a)}
    \label{fig:permIsTransl3}
\end{subfigure}
\caption{Demonstration of graph permutation as Euclidean translation. Figure \ref{fig:permIsTransl1} shows a signal lying on a lattice in the torus embedded onto a $5 \times 5$ planar grid. Figure \ref{fig:permIsTransl2} demonstrates a permutation of the graph and signal. Figure \ref{fig:permIsTransl3} shows a planar representation of the permuted graph and signal that is consistent with the one of Figure \ref{fig:permIsTransl1}. The permutation clearly corresponds to translations in a Euclidean space.}
\label{fig:permIsTransl}
\end{figure}

 {We first present two examples of permutations of graphs that can be viewed as Euclidean translations or rigid transformations. We later provide examples of permutations of the same graphs that are different than rigid transformations. The first example, demonstrated in Figure \ref{fig:permIsTransl1}, shows a periodic lattice graph $G$ and signal $\Bf$ with two values denoted by white and blue.} Note that the periodic graph can be embedded in a torus, whereas the figure only shows the projection of its 25 vertices into a $5 \times 5$ grid in the plane. The edges are not depicted in the figure, but they connect points to their four nearest neighbors on the torus. That is, including ``periodic padding'' for the $5 \times 5$ grid of vertices, each vertex in the plane is connected with its four nearest neighbors. For example vertex 21 is connected with vertices 1, 16, 22, 25. The graph signal obtains a non-zero constant value on the four vertices colored in blue (3, 4, 7 and 8) and is zero on the rest of them. Figure \ref{fig:permIsTransl2} demonstrates an application of a permutation $\BP$ to both the graph and the signal. At last, Figure \ref{fig:permIsTransl3} depicts the permuted graph and signal of Figure \ref{fig:permIsTransl2} when the indices are rearranged so that the representation of the lattice in the plane is the same as that in Figure \ref{fig:permIsTransl1} (this is necessary as the lattice lives in the torus and may have more than one representation in the plane). The relation between the consistent representations of ($G$,$\Bf$) in Figure \ref{fig:permIsTransl1} and ($\BP G$, $\BP \Bf$) in Figure \ref{fig:permIsTransl3} is obviously a translation. That is, graph and signal permutation in this example corresponds to translation. We remark that the fact that Figure \ref{fig:permIsTransl3} coincides with the description of ($G$,$\BP \Bf$) is incidental for this particular example and does not occur in the next example.

\begin{figure}[!ht]
\begin{tikzpicture}

\node[below=1.5cm, inner sep=0pt] at (0,0) {\parbox{0.3\linewidth}{\subcaption{$(G, \Bf)$
\label{fig:fivenodes1}}}};
\node[below=1.5cm, inner sep=0pt] at (5.5,3) {\parbox{0.3\linewidth}{\subcaption{$(\BP G, \BP \Bf)$
\label{fig:fivenodes2}}}};
\node[below=1.5cm, inner sep=0pt] at (11,0) {\parbox{0.3\linewidth}{\subcaption{$(\BP G, \BP \Bf)$ with indices of $\BP G$ embedded the same way as in (a) 
\label{fig:fivenodes3}}}};
\node[below=1.5cm, inner sep=0pt] at (5.5,-3.2) {\parbox{0.3\linewidth}{\subcaption{$(G, \BP \Bf)$
\label{fig:fivenodes4}}}};

\node[inner sep=0pt] (fivenodes1) at (0,0)
    {\includegraphics[width=0.3\linewidth]{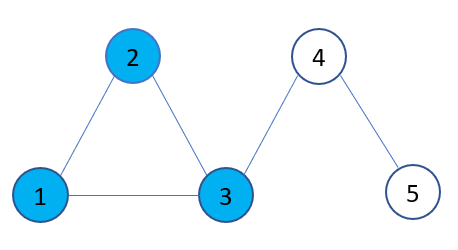}};
\node[inner sep=0pt] (fivenodes2) at (5.5,-3)
    {\includegraphics[width=0.32\linewidth]{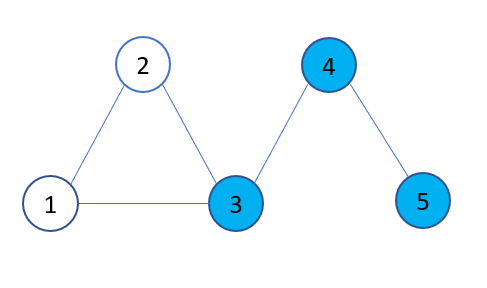}};
\node[inner sep=0pt] (fivenodes3) at (11,0)
    {\includegraphics[width=0.33\linewidth]{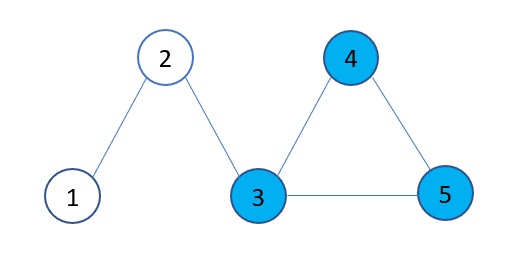}};
\node[inner sep=0pt] (fivenodes4) at (5.5,3)
    {\includegraphics[width=0.3\linewidth]{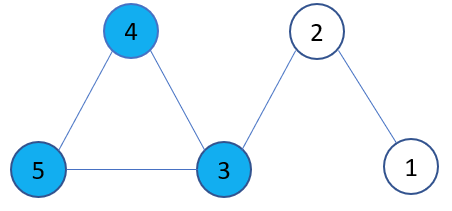}};

\draw[-] (0,2) -- (0,3);    
\draw[->] (0,3) -- (fivenodes4) node[midway, above] {relabel};
\draw[-] (fivenodes4) -- (11,3) node[midway, above] {rearrange};
\draw[->] (11,3) -- (11,2);
\end{tikzpicture}
\caption{Another example where a graph signal permutation corresponds to Euclidean translation. Figures \ref{fig:fivenodes1}-\ref{fig:fivenodes3} are created analogously to Figures \ref{fig:permIsTransl1}-\ref{fig:permIsTransl3}. Figure \ref{fig:fivenodes4} shows $(G, \BP \Bf)$, which is different from the rearrangement procedure depicted in \ref{fig:fivenodes3}.}
\label{fig:fivenodes}
\end{figure}
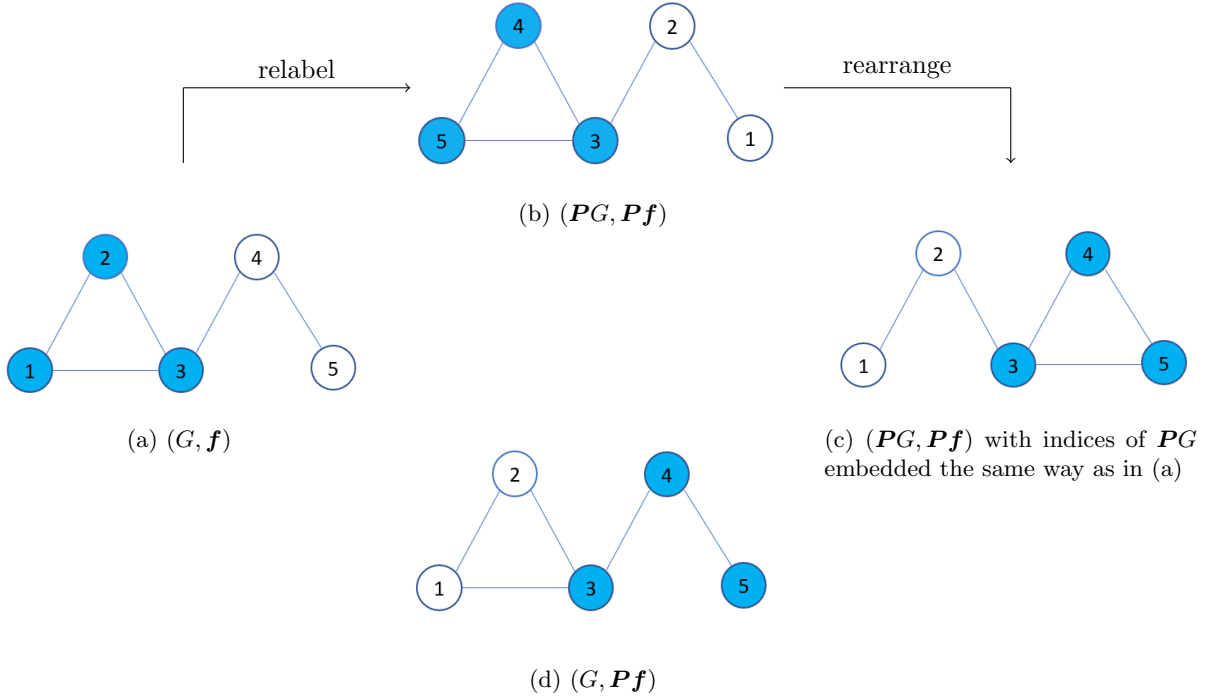

Figure \ref{fig:fivenodes} depicts a different example where a permutation of a graph signal can be viewed {as a variant of a Euclidean rigid transformation.}
The graph $G$ and the signal $\Bf$ are shown in Figure \ref{fig:fivenodes1}, where $\Bf$ is supported on the vertices marked in blue (indexed by 1, 2 and 3). Figure \ref{fig:fivenodes2} demonstrates an application of a permutation $\BP$ (mapping $(1,2,3,4,5)$ to $(5,4,3,2,1)$) to the graph and signal. Figure \ref{fig:fivenodes3} shows a different representation of $(\BP G, \BP \Bf)$, which is consistent with the one of $(G, \Bf)$ presented in Figure \ref{fig:fivenodes1}. The comparison between Figures \ref{fig:fivenodes1} and \ref{fig:fivenodes3} makes it clear that the graph and signal permutation corresponds to a Euclidean {rigid transformation} in the planar representation of the graph.  At last, Figure \ref{fig:fivenodes4} demonstrates that unlike the example in Figure \ref{fig:permIsTransl}, the rearrangement of $(\BP G, \BP \Bf)$
is generally different than 
the graph $(G, \BP \Bf)$. {Indeed, the subgraph associated with the blue values of the signal is not a triangle and thus the topology is different. }


\begin{figure}[!ht]
\centering
\begin{subfigure}{.25\textwidth}
	\centering
    \includegraphics[height=1.5in]{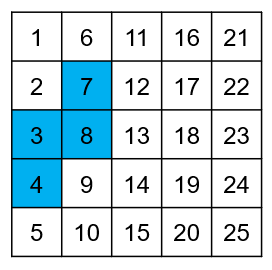}
    \caption{$(G, \Bf)$ \\ ~ }
    \label{fig:permIsTranslV1}
\end{subfigure}
{\large$\xrightarrow{\textup{~relabel~}}$}%
\begin{subfigure}{.25\textwidth}
	\centering
    \includegraphics[height=1.5in]{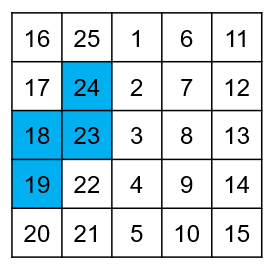}
    \caption{$(\BP G, \BP \Bf)$ \\ ~ }
    \label{fig:permIsTranslV2}
\end{subfigure}
{\large$\xrightarrow{\textup{rearrange}}$}%
\begin{subfigure}{.25\textwidth}
	\centering
    \includegraphics[height=1.5in]{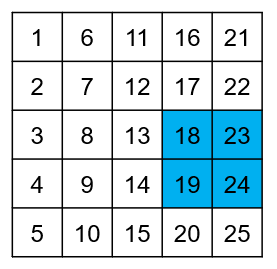}
    \caption{Indices of $(\BP G, \BP \Bf)$ \\ rearranged as in (a)}
    \label{fig:permIsTranslV3}
\end{subfigure}
\caption{{A different permutation of Figure \ref{fig:permIsTransl1}, which is not similar to rigid motion, but still preserves the graph topology. Note that \ref{fig:permIsTranslV3} does not maintain the planar geometry of the graph: for instance, the vertices 19 and 24 are not connected by an edge. }}
\label{fig:permIsTranslV}
\end{figure}

{We remark that many permutations on graphs do not act like translations or rigid transformations. We demonstrate this claim using the graphs of the previous two examples. In Figure \ref{fig:permIsTranslV}, we consider the same graph as in Figure \ref{fig:permIsTransl}, but with a different permutation. The difference in permutations can be noticed by comparing the second columns of the grids in Figures \ref{fig:permIsTransl2} and \ref{fig:permIsTranslV2}. We note that the rearrangement of the vertices in Figure \ref{fig:permIsTranslV3} does not yield an analog of a Euclidean translation. The reason is that the rearranged  vertices do not form a grid. To demonstrate this claim, note that in Figure \ref{fig:permIsTranslV1}, label 17 is connected to 22, but in Figure \ref{fig:permIsTranslV2}, and consequently in the rearranged representation in Figure \ref{fig:permIsTranslV3}, they are disconnected.}

\begin{figure}[!ht]
\centering
\begin{subfigure}{.25\textwidth}
	\centering
    \includegraphics[height=.8in]{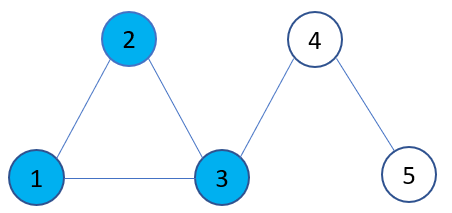}
    \caption{$(G, \Bf)$ \\ ~ }
    \label{fig:fivenodesV1}
\end{subfigure}
{\large$\xrightarrow{\textup{~relabel~}}$}%
\begin{subfigure}{.25\textwidth}
	\centering
    \includegraphics[height=.8in]{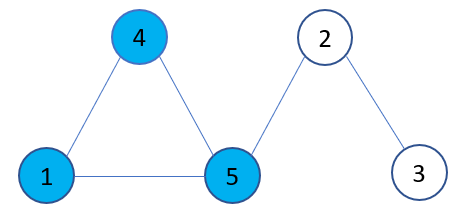}
    \caption{$(\BP G, \BP \Bf)$ \\ ~ }
    \label{fig:fivenodesV2}
\end{subfigure}
{\large$\xrightarrow{\textup{rearrange}}$}%
\begin{subfigure}{.25\textwidth}
	\centering
    \includegraphics[height=.8in]{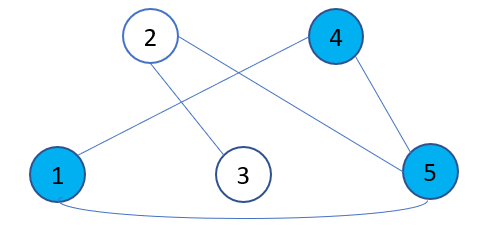}
    \caption{Indices of $(\BP G, \BP \Bf)$ \\ rearranged as in (a)}
    \label{fig:fivenodesV3}
\end{subfigure}
\caption{{A different permuation of Figure \ref{fig:fivenodes1}, which is not similar to rigid motion, but still preserves the graph topology.}}
\label{fig:fivenodesV}
\end{figure}

{Figure \ref{fig:fivenodesV} demonstrates a permutation that does not act like a rigid transformation with respect to the graph of Figure \ref{fig:fivenodes}. Clearly, the rearranged graph and signal in Figure \ref{fig:fivenodesV3} have different planar geometry. We remark that while the permutations demonstrated in Figures \ref{fig:permIsTranslV} and \ref{fig:fivenodesV} do not preserve the planar geometry, they still preserve the topology of the graphs. Indeed, the notion of permutation invariance is richer than invariance to rigid transformations in the Euclidean domain.}

In the signal processing community, {some} candidates were proposed for translating signals on graphs. Shuman et al.~\cite{ShuRV16} {defined a ``graph translation'' (or in retrospect, a graph localization procedure) as follows} 
\begin{equation}\label{eq:defTranslMod}
T_c \Bf = \sqrt{N} (\Bf \ast \boldsymbol{\delta}_c) = \sqrt{N} \sum_{l=0}^{N-1} u_l(c) \Bu_l \Bu_l^* \Bf  ~.
\end{equation}
{They established useful localization properties of $T_c$, which justify a corresponding construction of a windowed graph Fourier transform. They also demonstrated the applicability of this tool for the Minnesota road network \cite[Figure 7]{ShuRV16}. 
We remark that in their definition $u_l(c)$ may not be well-defined. To make it well-defined one needs to assume fixed choices of the phases of $u_l(c)$, $0 \leq l \leq N-1$, and that the algebraic multiplicities of all eigenvalues equal one. }

Sandryhaila and Moura \cite{SanM13} define a ``shift'' of a graph signal $\Bf$ by $T_s \Bf = \BW \Bf$, where $\BW$ is the weight matrix of the graph. This definition is motivated by the example of a directed cyclic graph, where an application of the weight matrix is equivalent to a shift by one vertex. Note that in this special case, the graph signal permutation $(\BP G, \BP \Bf)$ advocated in this section also results in a vertex shift. {We remark that it is unclear to us why this notion of shift is useful for general graphs.}

If one needs covariance and approximate invariance of a graph scattering transform to the {graph localization procedure} defined in \cite{ShuRV16}, then one may modify the nonlinearity of the scattering transform as $\sigma(\Bf) = \sum_{l=0}^{N-1} \abs{\Bu_l^* \Bf} \Bu_l$ and redefine $\BU_j \Bf = \sigma(\BQ_j \Bf)$ for $j > -J$. Note that
\begin{equation}
\begin{aligned}
\sigma(T_c \Bf) ~=~  \sigma \left( \sqrt{N} \sum_{l=0}^{N-1} u_l(c) (\Bu_l^* \Bf) \Bu_l \right)
~=~  \sqrt{N} \sum_{l=0}^{N-1} u_l(c) \abs{\Bu_l^* \Bf} \Bu_l
\end{aligned}
\end{equation}
and
\begin{equation}
\begin{aligned}
T_c \sigma(\Bf) ~=~  \sigma \sum_{l=0}^{N-1} u_l(c) \left( \Bu_l^* \sum_{l'=0}^{N-1}  \abs{\Bu_{l'}^* \Bf} \Bu_{l'} \right) \Bu_l
~=~  \sqrt{N} \sum_{l=0}^{N-1} u_l(c) \abs{\Bu_l^* \Bf} \Bu_l ~.
\end{aligned}
\end{equation}
Therefore, the nonlinearity $\sigma$ and the modified scattering transform are covariant to the {localization operator} $T_c$. Similarly, by following the proof for Theorem \ref{thm:transinv} one can show that the modified scattering transform is approximately invariant to {$T_c$} as long as its energy decay is sufficiently fast.

The scattering transform cannot be adjusted to be covariant and approximately invariant to the ``shift'' defined by $T_s \Bf =  \BW \Bf$. The reason is that unlike $\BL$, $\BW$ does not commute in general with the eigenvectors $\{\Bu_l\}_{l=1}^N$.

\subsection{Proof of Proposition \ref{prop:permCov}}\label{subsec:proofpermcov}
We need to show that for each path $p = (j_1, \cdots, j_m) \in \cP_J$, where $j_1, \cdots, j_m > -J$, 
\begin{equation}\label{eq:needForEachPath}
\BS[\BP G] [p] \BP \Bf = \BP \BS [G] [p] \Bf ~.
\end{equation}
Note that the Laplacian of $\BP G$ is $\tilde{\BL} = \BP \BL \BP^*$, which has the same eigenvalues as $\BL$ and has eigenvectors $\tilde{\Bu}_l =  \BP \Bu_l$, $l = 0, \cdots, N-1$. Equation (\ref{eq:defconvpsi}) implies that for $j > -J$
\begin{equation}
\Bf \ast_{\BP G} \boldsymbol{\psi}_j = \sum_{l=0}^{N-1} \BP \Bu_l \Bu_l^* \BP^* \Bf \hat{\psi} (2^{-j} \l_l) ~.
\end{equation}
Therefore, for $j > -J$
\begin{equation}
\begin{aligned}
(\BP \Bf) \ast_{\BP G} \boldsymbol{\psi}_j ~=~ & \sum_{l=0}^{N-1} \BP \Bu_l \Bu_l^* \BP^* \BP \Bf \hat{\psi} (2^{-j} \l_l) \\
~=~ & \sum_{l=0}^{N-1} \BP \Bu_l \Bu_l^* \Bf \hat{\psi} (2^{-j} \l_l) = \BP \sum_{l=0}^{N-1}  \Bu_l \Bu_l^* \Bf \hat{\psi} (2^{-j} \l_l) = \BP (\Bf \ast_{G} \boldsymbol{\psi}_j).
\end{aligned}
\end{equation}
Consequently, applying the absolute value pointwise,
\begin{equation}\label{eq:PfconvPG}
\abs{(\BP \Bf) \ast_{\BP G} \boldsymbol{\psi}_j} = \abs{\BP (\Bf \ast_{G} \boldsymbol{\psi}_j)} = \BP \abs{\Bf \ast_G \boldsymbol{\psi}_j}
\end{equation}

Similarly,
\begin{equation}\label{eq:PfconvG}
\abs{(\BP \Bf) \ast_{\BP G} \boldsymbol{\phi}_{-J}} = \BP \abs{ (\Bf \ast_{G} \boldsymbol{\phi}_{-J}) } ~.
\end{equation}

Application of (\ref{eq:PfconvPG}) and (\ref{eq:PfconvG}) results in the identity
\begin{equation}\label{eq:fancyExpression}
\Big | \abs{(\BP \Bf) \ast_{\BP G} \boldsymbol{\psi}_{j_1}} \ast_{\BP G} \cdots \ast_{\BP G} \boldsymbol{\psi}_{j_m} \Big | \ast_{\BP G} \boldsymbol{\phi}_{-J} = \BP \Big | \abs{ \Bf \ast_{G} \boldsymbol{\psi}_{j_1}} \ast_{G} \cdots \ast_{G} \boldsymbol{\psi}_{j_m} \Big | \ast_{G} \boldsymbol{\phi}_{-J} ~.
\end{equation}
In view of (\ref{eq:defUp}) -- (\ref{eq:defSf}),  (\ref{eq:needForEachPath}) is equivalent to (\ref{eq:fancyExpression}), and the proof is thus concluded.

\subsection{Proof of Theorem \ref{thm:transinv}}\label{subsec:prooftransinv}
According to (\ref{eq:defSf}) and (\ref{eq:permCov}),
\begin{equation}\label{eq:opratorSplit}
\begin{aligned}
\norm{\BS[\BP G][\cP_J] \BP \Bf - \BS[G][\cP_J] \Bf} ~=~ & \norm{\BP \BQ_J \BU[G][\cP_J] \Bf - \BQ_J \BU[G][\cP_J] \Bf} \\
~\leq~ & \norm{\BP \BQ_J -  \BQ_J} \norm{\BU[\cP] \Bf} ~.
\end{aligned}
\end{equation}

We bound the right-hand-side of (\ref{eq:opratorSplit}) by a function that approaches zero as $J \rightarrow \infty$. We first bound $\norm{\BP \BQ_J -  \BQ_J}$. 
We apply (\ref{eq:defSf}) as well as the following facts: $\l_0 = 0$, $\hat{\phi}(0) = 0$ and $\l_1 > 0$ (since $G$ is connected) to obtain that for $\Bf \in \CC^N$
\begin{equation}
\begin{aligned}
\norm{(\BP \BQ_J - \BQ_J) \Bf}^2 ~=~ & \norm{\BP \sum_{l=0}^{N-1} \hat{\phi}(2^J \l_l) \Bu_l \Bu_l^* \Bf - \sum_{l=0}^{N-1} \hat{\phi}(2^J \l_l) \Bu_l \Bu_l^* \Bf}^2 \\
~=~ & \norm{\sum_{l=0}^{N-1} \hat{\phi}(2^J \l_l) \Bu_l \Bu_l^* (\BP\Bf-\Bf)}^2 \\
~=~ & \norm{\sum_{l=1}^{N-1} \hat{\phi}(2^J \l_l) \Bu_l \Bu_l^* (\BP\Bf-\Bf)}^2 \\
~=~ & \sum_{l=1}^{N-1} \abs{\hat{\phi}(2^J \l_l) \Bu_l^* (\BP\Bf-\Bf)}^2 \\
~\leq~ & \max_{l=1, \cdots, N-1} \abs{\hat{\phi}(2^J \l_l)}^2 \norm{\BP-\BI}^2 \norm{\Bf}^2 \\
~\leq~ & C_{\phi}^2 2^{-2J} \l_1^{-2} \norm{\BP-\BI}^2 \norm{\Bf}^2 ~.
\end{aligned}
\end{equation}
Hence
\begin{equation}
\label{eq:assist2_app_inv}
\norm{\BP \BQ_J - \BQ_J} \leq C_{\phi} 2^{-J} \l_1^{-1} \norm{\BP-\BI} ~.
\end{equation}
It remains to bound $\norm{\BU[\cP_J] \Bf}$. 
The application of (\ref{eq:defUPSP}), Proposition \ref{prop:layer} and (\ref{eq:fSingleSplit}) results in 
\begin{equation}
\label{eq:assist3_app_inv}
\begin{aligned}
\norm{\BU[\cP_J] \Bf}^2 ~=~ & \norm{\Bf}^2 + \sum_{m \geq 1} \sum_{p \in \Lambda^m} \norm{\BU[p] \Bf}^2 \\
~\leq~ & \norm{\Bf}^2 + \sum_{m \geq 1} \left( 1 - \frac{2}{N} \right)^{m-1} \sum_{p \in \Lambda^1} \norm{\BU[p] \Bf}^2 \\
~=~ & \norm{\Bf}^2 + \frac{N}{2} \sum_{p \in \Lambda^1} \norm{\BU[p] \Bf}^2 \\
~\leq~ & \norm{\Bf}^2 + \frac{N}{2} \norm{\Bf}^2 \\
~=~ & \frac{N+2}{2} \norm{\Bf}^2 ~.
\end{aligned}
\end{equation}
At last, the combination of \eqref{eq:opratorSplit}, \eqref{eq:assist2_app_inv} and \eqref{eq:assist3_app_inv} implies \eqref{eq:approx_inv}. The generalization to $\BF \in \CC^{N \times D}$ is immediate since $\BF = (\Bf_1, \cdots, \Bf_D)$ and $\norm{\BF}_{\F}^2 = \sum_{d=1}^D \norm{\Bf_d}^2$.

\section{Stability to {signal and} graph manipulations}\label{sec:stability}
{We establish the stability of the graph scattering transform to both signal and graph manipulations. The stability to signal manipulation is an immediate corrolary of the energy preservation established in Theorem \ref{thm:normPreserving}. It states that the graph scattering transform is Lipschitz continuous with respect to the graph signal in the following way.}
{
\begin{proposition}
\label{cor:stabilitysignal}
For two signals $\Bf \in \CC^N$ and $\tilde{\Bf} \in \CC^N$,
\begin{equation}
\label{eq:prop5.1_1}
    \norm{ \BS[\cP_J] \Bf - \BS[\cP_J] \tilde{\Bf} } \leq \norm{ \Bf - \tilde{\Bf} } ~.
\end{equation}
Similarly, for two signals $\BF \in \CC^{N \times D}$ and $\tilde{\BF} \in \CC^{N \times D}$,
\begin{equation}
\label{eq:prop5.1_2}
    \norm{ \BS[\cP_J] \BF - \BS[\cP_J] \tilde{\BF} }_{\F} \leq \norm{ \BF - \tilde{\BF} }_{\F} ~.
\end{equation}
\end{proposition}
}

{In order to motivate the stability to graph manipulation, we discuss the problem of community detection \cite{For10}. Its setting assumes different groups of vertices that communicate more significantly with each other than with other groups. The goal is to identify these underlying groups.} In some cases, such as for data generated by the stochastic-block model \cite{Abb17}, the edge set of the graph is the only information one can work with. For other cases, such as bibliographic datasets \cite{SenNB08}, in addition to the edge set (the citations), information of features of vertices is provided. For graph convolutional neural networks, if vertex-wise features are not given, it is natural to choose an artificial feature for each vertex. For instance, Kipf \& Welling \cite{KipW16} use $\Bf = (1,1,\cdots,1)^*$ and Bruna \& Li \cite{BruL17} use $\BF = \BI$.

{In this problem, stability to graph manipulations can be formulated as follows when the number of vertices is sufficiently large: small changes of the edge weights should not affect the community structure.
More specifically, one may consider graph manipulation as modification of edge weights and ask for the effect on such manipulation on important features. In the following, we establish such stability to graph manipulations, where the features are expressed by the output of the graph scattering transform. This result is conditioned on sufficiently fast decay rate of the energy as well as of $\phi$ and $\psi$ (equivalently, their Fourier transforms are sufficiently smooth).
\begin{theorem}\label{thm:weightperturb}
Let $G$ be a simple graph with $N$ vertices and weights $\{\BW(n,m)\}_{n,m=1}^N$, and let $\delta > 0$ denote the smallest gap of eigenvalues of its Laplacian:
\begin{equation*}
    \delta = \min_{{l_1} \neq {l_2}} \abs{\l_{l_1} - \l_{l_2}} ~.
\end{equation*}
Let $\tilde{G}$ be a perturbation of $G$ with weights $\{\tilde{\BW}(n,m)\}_{n,m=1}^N$, such that for some $0 < C_{\sharp} \leq N \delta/2$
\begin{equation}
\label{eq:weak_cond_in_thm}
    \abs{\BW(n,m) - \tilde{\BW}(n,m)} \leq C_{\sharp} N^{-2} ~.
\end{equation}
Let $\Bf \in \CC^N$ be a fixed input signal for which the energy of the scattering transform decays fast in the sense that for some $M > 0$ and $C_0 > 0$ 
\begin{equation}
\sum_{m \geq M} \sum_{p \in \Lambda^m} \norm{\BS[p] \Bf}^2 \leq \frac{C_0}{N} \norm{\Bf}^2 ~.
\end{equation}
Also, suppose $\hat{\phi}$ and $\hat{\psi}$ are both Lipschitz continuous functions with Lipschitz constant $C_1$.
Then there exists a constant C depending on $C_{\sharp}$, $C_0$, $C_1$, such that
\begin{equation}
\norm{\BS[G][\cP_J]\Bf-\BS[\tG][\cP_J]\Bf} \leq \frac{C}{\sqrt{N}} \norm{\Bf} ~.
\end{equation}
The same result holds if $\Bf \in \CC^N$ is replaced with $\BF \in \CC^{N \times D}$.
\end{theorem}
}

{We remark that the assumption $\delta > 0$ in the above theorem implies that all eigenvalues have algebraic multiplicity one. In general, it is impossible to extend this theorem to higher multiplicity of eigenvalues. Indeed, assume for example that there are two zero eigenvalues, so the graph is disconnected. Then it is possible to make the graph connected by changing a certain edge weight from zero to an arbitrarily small positive number. Such a small change completely deform the topology of the graph and we thus do not expect a general theorem that includes higher multiplicities.}

{We also remark that \eqref{eq:weak_cond_in_thm} only allows a very small change of weights and generally does not allow one to remove or add an edge.} {The latter more general graph manipulation is natural in some applications. For example, for the CORA dataset \cite{SenNB08} of publications and citations, the lack of knowledge of the mutual citation between two specific publications should not significantly affect the detection of communities. We are unaware of previous theoretical results for stability with respect to this more general graph manipulation.}
{In some cases, removing an edge from a graph can completely change the topology, no matter how large the graph is. For example, one can make some graphs disconnected by removing a single edge. Therefore, it is difficult to have a general result that can handle stability to removal or addition of edges. Nevertheless, the following theorem generalizes Theorem \ref{thm:weightperturb} by restricting the perturbation of the spectral decomposition of the graph Laplacian.}

\begin{theorem}\label{thm:forgeneral}
Let $G$ and $\tG$ be two simple graphs with the same set of $N$ vertices. Let $\Bf \in \CC^N$ be a fixed input signal for which the energy of the scattering transform decays fast in the sense that for some $M > 0$ and $C_0 > 0$ 
\begin{equation}
\sum_{m \geq M} \sum_{p \in \Lambda^m} \norm{\BS[p] \Bf}^2 \leq \frac{C_0}{N} \norm{\Bf}^2 ~.
\end{equation}
Also, suppose $\hat{\phi}$ and $\hat{\psi}$ are both Lipschitz continuous functions with Lipschitz constant $C_1$ and the Laplacian eigenpairs of $G$ and $\tG$ satisfy 
\begin{equation}\label{eq:eigenConditions}
\abs{\l_l - \tl_l} \leq \frac{C_2}{N} \quad and \quad \sin \angle (\Bu_l, \tilde{\Bu}_l) \leq \frac{C_3}{N},  \quad l = 0, \cdots, N-1 ~.
\end{equation}
Then there exists a constant $C$ depending on $C_0$, $C_1$, $C_2$, $C_3$ and $M$, for which
\begin{equation}\label{eq:stablegraph}
\norm{\BS[G][\cP_J]\Bf-\BS[\tG][\cP_J]\Bf} \leq \frac{C}{\sqrt{N}} \norm{\Bf} ~.
\end{equation}
The same result holds if $\Bf \in \CC^N$ is replaced with $\BF \in \CC^{N \times D}$.
\end{theorem}

\begin{figure}[!ht]
    \centering
    \includegraphics[width=.5\linewidth]{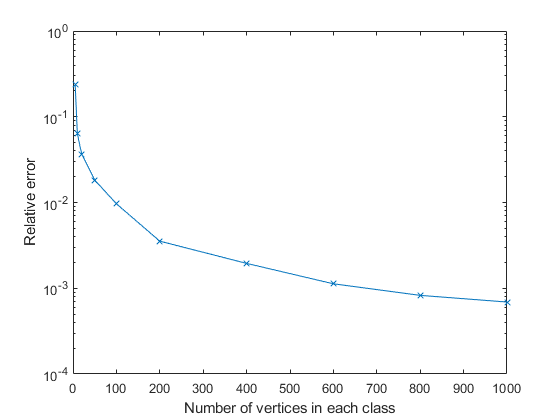}
	\caption{The relative error $|| \BS[G]\Bf-\BS[\tG]\Bf || / \norm{\Bf} $ for a graph $G$ and its perturbation $\tilde{G}$ generated by SBM. The perturbation $\tilde{G}$
is formed by randomly selecting two vertices of $G$ and reversing the connectivity by adding/deleting an edge between them.
The average ratio from 20 randomly generated graphs is plotted for each $N$ (number of vertices in each class).}
    \label{fig:sbmStability}
\end{figure}

Condition (\ref{eq:eigenConditions}) might be strong for some practical applications, but we are unable to relax it. {As mentioned above, we do not expect a general stability result with respect to addition or removal of edges. Nevertheless, for some stochastic models, it is rather unlikely that adding or removing an edge leads to a significant change in the graph topology. To demonstrate this claim, we numerically test} the stability of the scattering transform in practice for a synthetic setting produced by SBM with arbitrary edge corruption. In this setting, the condition of \eqref{eq:eigenConditions} may not hold. For each $N = 5,10,20,50,100,200,400,600,800,1,000$, we randomly sample 20 graphs from an SBM with two classes both containing $N$ vertices. The probability of connecting two vertices within one class is $p = \max \{1, 6 \log N / N\}$ and the probability of connecting two vertices from different classes is $q = \log N / N$. 
For each model, 
we randomly choose two vertices of the graph $G$: if they are connected by an edge, we remove the edge to form $\tilde{G}$; if they are not connected by an edge, we add an edge to form $\tilde{G}$. 
We compute the relative error $\norm{\BS[G]\Bf-\BS[\tG]\Bf} / \norm{\Bf}$  and average it over the 20 random samples. The results of the experiments are shown in Figure \ref{fig:sbmStability}. We note that the ratio decays fast and the relative error is negligible for sufficiently large graphs.

\subsection{Proof of Proposition \ref{cor:stabilitysignal}}

{
Similarly to establishing \eqref{eq:pathbreakeq},
\begin{align}
    \sum_{p \in \Lambda^m} \norm{ \BU[p]\Bf - \BU[p]\tilde{\Bf} }^2 ~=~ & \sum_{p \in \Lambda^m} \left( \sum_{j > -J} \norm{ (\BU[p] \Bf - \BU[p] \tilde{\Bf}) \ast \Bvpsi_j }^2 + \norm{ (\BU[p] \Bf - \BU[p] \tilde{\Bf}) \ast \Bvphi_J }^2 \right) \nonumber \\ \nonumber
    ~=~ & \sum_{p \in \Lambda^m} \left( \sum_{j > -J} \norm{ \BU[p] \Bf \ast \Bvpsi_j - \BU[p] \tilde{\Bf} \ast \Bvpsi_j }^2 + \norm{ \BU[p] \Bf \ast \Bvphi_J - \BU[p] \tilde{\Bf} \ast \Bvphi_J }^2 \right) \\
    ~\geq~ & \sum_{p \in \Lambda^m} \sum_{j > -J} \norm{\BU[p+j]\Bf - \BU[p+j]\tilde{\Bf}}^2 + \sum_{p \in \Lambda^m} \norm{\BS[p]\Bf - \BS[p]\tilde{\Bf}}^2  \label{eq:clarify_u_dif} \\ \nonumber
    ~=~ & \sum_{p \in \Lambda^{m+1}} \norm{\BU[p]\Bf - \BU[p]\tilde{\Bf}}^2 + \sum_{p \in \Lambda^m} \norm{\BS[p]\Bf - \BS[p]\tilde{\Bf}}^2 ~.
\end{align}
We remark that the inequality of \eqref{eq:clarify_u_dif} follows from the triangle inequality $\norm{\Bx-\By} \geq \abs{ \norm{\Bx} - \norm{\By} }$. 
By summing the terms on the left and right hand sides of \eqref{eq:clarify_u_dif}  over $m \geq 0$, one obtains, similarly to \eqref{eq:assist1_normPreserving}, that 
\begin{equation}
\label{eq:assist_proof_prop5.1_2}
    \norm{\BS[\cP_J]\Bf - \BS[\cP_J]\tilde{\Bf}}^2 ~\leq~ \norm{\Bf - \tilde{\Bf}}^2 - \lim_{m \rightarrow \infty} \sum_{p \in \Lambda^{m+1}} \norm{\BU[p]\Bf - \BU[p]\tilde{\Bf}}^2~.
\end{equation}
Application of \eqref{eq:assist2_normPreserving} to \eqref{eq:assist_proof_prop5.1_2} yields \eqref{eq:prop5.1_1}. The proof of \eqref{eq:prop5.1_2} is the same.
}

\subsection{Proof of Theorem \ref{thm:forgeneral}}
Let $p \in \Lambda^m$ be an arbitrary path of length $m \geq 0$. We bound the squared norm of the difference of wavelet coefficients of $\BU[p] \Bf$ with respect to $G$ and $\tG$. Recall that $\l_0 = \tl_0 = 0$ and $\Bu_0 = \tilde{\Bu}_0 = (1/\sqrt{N}, \cdots, 1/\sqrt{N})^*$. The required bound for the $J$-th coefficient is as follows
\begin{equation}\label{eq:onestepJ}
\begin{aligned}
& \norm{\BQ_J[G]\BU[G][p]\Bf - \BQ_J[\tG]\BU[\tG][p]\Bf}^2 \\
~=~ & \norm{\sum_{l=0}^{N-1} \hat{\phi}(2^J \l_l) \Bu_l \Bu_l^* \BU[G][p]\Bf - \sum_{l=0}^{N-1} \hat{\phi}(2^J \tl_l) \Btu_l \Btu_l^* \BU[\tG][p]\Bf}^2 \\
~=~ & \Bigg\Vert \sum_{l=0}^{N-1} \left( \hat{\phi}(2^J \l_l)-\hat{\phi}(2^J \tl_l) \right) \Bu_l \Bu_l^* \BU[G][p]\Bf  +  \sum_{l=0}^{N-1} \hat{\phi}(2^J \l_l) \left( \Bu_l \Bu_l^* - \Btu_l \Btu_l^* \right) \BU[G][p]\Bf  + \\
& \qquad \sum_{l=0}^{N-1} \hat{\phi}(2^J \tl_l) \Btu_l \Btu_l^* \left( \BU[G][p]\Bf - \BU[\tG][p]\Bf \right) \Bigg\Vert^2 \\
~=~ & \Bigg\Vert \sum_{l=1}^{N-1} \left( \hat{\phi}(2^J \l_l)-\hat{\phi}(2^J \tl_l) \right) \Bu_l \Bu_l^* \BU[G][p]\Bf  +  \sum_{l=1}^{N-1} \hat{\phi}(2^J \l_l) \left( \Bu_l \Bu_l^* - \Btu_l \Btu_l^* \right) \BU[G][p]\Bf  + \\
& \qquad \sum_{l=0}^{N-1} \hat{\phi}(2^J \tl_l) \Btu_l \Btu_l^* \left( \BU[G][p]f - \BU[\tG][p]\Bf \right) \Bigg\Vert^2 \\
~\leq~ & 3 \Bigg( \norm{\sum_{l=1}^{N-1} \left( \hat{\phi}(2^J \l_l)-\hat{\phi}(2^J \tl_l) \right) \Bu_l \Bu_l^* \BU[G][p]\Bf}^2 + \norm{\sum_{l=1}^{N-1} \hat{\phi}(2^J \l_l) \left( \Bu_l \Bu_l^* - \Btu_l \Btu_l^* \right) \BU[G][p]\Bf}^2 + \\
& \qquad \norm{\sum_{l=0}^{N-1} \hat{\phi}(2^J \tl_l) \Btu_l \Btu_l^* \left( \BU[G][p]\Bf - \BU[\tG][p]\Bf \right)}^2 \Bigg) ~.
\end{aligned}
\end{equation}
Similarly, for $j > -J$, 
\begin{equation}\label{eq:onestepj}
\begin{aligned}
& \norm{\BQ_j[G]\BU[G][p]\Bf - \BQ_j[\tG]\BU[\tG][p]\Bf}^2 \\
~\leq~ & 3 \Bigg( \norm{\sum_{l=1}^{N-1} \left( \hat{\psi}(2^{-j} \l_l)-\hat{\psi}(2^{-j} \tl_l) \right) \Bu_l \Bu_l^* \BU[G][p]\Bf}^2 + \norm{\sum_{l=1}^{N-1} \hat{\psi}(2^{-j} \l_l) \left( \Bu_l \Bu_l^* - \Btu_l \Btu_l^* \right) \BU[G][p]\Bf}^2 + \\
& \qquad \norm{\sum_{l=0}^{N-1} \hat{\psi}(2^{-j} \tl_l) \Btu_l \Btu_l^* \left( \BU[G][p]\Bf - \BU[\tG][p]\Bf \right)}^2 \Bigg) ~.
\end{aligned}
\end{equation}

Both (\ref{eq:onestepJ}) and (\ref{eq:onestepj}) bound the energy by the sum of three terms. We next bound each of these terms. In order the bound the first term, note that since $\hat{\phi}$ and $\hat{\psi}$ are both $C_1$-Lipschitz
\begin{equation}
\abs{\hat{\phi}(2^J \l_l)-\hat{\phi}(2^J \tl_l)} \leq C_1 2^J \abs{\l_l - \tl_l} ~,
\end{equation}
and
\begin{equation}
\abs{\hat{\psi}(2^{-j} \l_l)-\hat{\psi}(2^{-j} \tl_l)} \leq C_1 2^{-j} \abs{\l_l - \tl_l} ~, \quad \mbox{for all }~ j > -J ~.
\end{equation}
Therefore,
\begin{equation}\label{eq:firstUsefulBound}
\begin{aligned}
& \norm{\sum_{l=1}^{N-1} \left( \hat{\phi}(2^J \l_l)-\hat{\phi}(2^J \tl_l) \right) \Bu_l \Bu_l^* \BU[G][p]\Bf}^2 \\
~\leq~ & \max_{l = 1, \cdots, N-1} \abs{\hat{\phi}(2^J \l_l) - \hat{\phi}(2^J \tl_l)}^2 \sum_{l=1}^{N-1} \norm{\Bu_l \Bu_l^* \BU[G][p]\Bf}^2 \\
~\leq~ & C_1 2^{2J} \frac{C_2}{N} \norm{\BU[G][p]\Bf}^2 \\
~=~ & \frac{C_1 C_2 2^{2J}}{N} \norm{\BU[G][p]\Bf}^2 ~.
\end{aligned}
\end{equation}
Similarly,
\begin{equation}
\norm{\sum_{l=1}^{N-1} \left( \hat{\psi}(2^{-j} \l_l)-\hat{\psi}(2^{-j} \tl_l) \right) \Bu_l \Bu_l^* \BU[G][p]\Bf}^2 \leq \frac{C_1 C_2 2^{-2j}}{N} \norm{\BU[G][p]\Bf}^2 ~.
\end{equation}

The second term for $J$ and $j > -J$ is bounded as follows
\begin{equation}
\begin{aligned}
& \norm{\sum_{l=1}^{N-1} \hat{\phi}(2^J \l_l) (\Bu_l \Bu_l^* - \Btu_l \Btu_l^*) \BU[G][p]\Bf}^2 \\
~\leq~ & \sum_{l=0}^{N-1} \norm{\Bu_l \Bu_l^* - \Btu_l \Btu_l^*}^2 \abs{\hat{\phi}(2^J \l_l)}^2 \norm{\BU[G][p]\Bf}^2 \\
~\leq~ & \frac{C_3^2}{N^2} \sum_{l=1}^{N-1} \abs{\hat{\phi}(2^J \l_l)}^2 \norm{\BU[G][p]\Bf}^2 
\end{aligned}
\end{equation}
and
\begin{equation}
\begin{aligned}
\norm{\sum_{l=1}^{N-1} \hat{\psi}(2^{-j} \l_l) (\Bu_l \Bu_l^* - \Btu_l \Btu_l^*) \BU[G][p]\Bf}^2
\leq & \frac{C_3^2}{N^2} \sum_{l=1}^{N-1} \abs{\hat{\psi}(2^{-j} \l_l)}^2 \norm{\BU[G][p]\Bf}^2 ~.
\end{aligned}
\end{equation}

The third term for $J$ and $j > -J$ has the form
\begin{equation}
\begin{aligned}
\norm{\sum_{l=0}^{N-1} \hat{\phi}(2^J \tl_l) \Btu_l \Btu_l^* \left( \BU[G][p]\Bf - \BU[\tG][p]\Bf \right)}^2
= & \sum_{l=0}^{N-1} \abs{\hat{\phi}(2^J \tl_l)}^2 \norm{\Btu_l^* (\BU[G][p]f - \BU[\tG][p]\Bf)}^2 
\end{aligned}
\end{equation}
and
\begin{equation}\label{eq:lastUsefulBound}
\begin{aligned}
\norm{\sum_{l=0}^{N-1} \hat{\psi}(2^{-j} \tl_l) \Btu_l \Btu_l^* \left( \BU[G][p]\Bf - \BU[\tG][p]\Bf \right)}^2
= & \sum_{l=0}^{N-1} \abs{\hat{\psi}(2^{-j} \tl_l)}^2 \norm{\Btu_l^* (\BU[G][p]\Bf - \BU[\tG][p]\Bf)}^2 ~.
\end{aligned}
\end{equation}

Applying (\ref{eq:firstUsefulBound}) -- (\ref{eq:lastUsefulBound}), 
\begin{equation}
\begin{aligned}
& \norm{\BQ_J[G]\BU[G][p]\Bf - \BQ_J[\tG]\BU[\tG][p]\Bf}^2 + \sum_{j > -J} \norm{\BQ_j[G]\BU[G][p]\Bf - \BQ_j[\tG]\BU[\tG][p]\Bf}^2 \\
~\leq~ & 3 \Bigg( \sum_{j \geq -J} \frac{C_1 C_2 2^{-2j}}{N} \norm{\BU[G][p]\Bf}^2 + \frac{C_3^2}{N^2} \sum_{l=1}^{N-1} \bigg( \abs{\hat{\phi}(2^J \l_l)}^2 + \sum_{j > -J} \abs{\hat{\psi}(2^{-j}\l_l)}^2 \bigg) \norm{\BU[G][p]\Bf}^2 + \\
& \qquad \sum_{l=0}^{N-1} \bigg( \abs{\hat{\phi}(2^J \l_l)}^2 + \sum_{j > -J} \abs{\hat{\psi}(2^{-j}\l_l)}^2 \bigg) \norm{\Btu_l^* \left( \BU[G][p]\Bf - \BU[\tG][p]\Bf \right)}^2 \Bigg) \\
~=~ & 3 \Bigg( \frac{C_1 C_2 2^{2J+1}}{N} \norm{\BU[G][p]\Bf}^2 + \frac{(N-1)C_3^2}{N^2} \norm{\BU[G][p]\Bf}^2 + \norm{\BU[G][p]f-\BU[\tG][p]\Bf}^2 \Bigg) \\
~\leq~ & 3 \Bigg( \frac{C}{N} \norm{\BU[G][p]\Bf}^2 + \norm{\BU[G][p]\Bf-\BU[\tG][p]\Bf}^2 \Bigg) ~,
\end{aligned}
\end{equation}
where $C = C_1 C_2 2^{2J+1} + C_3^2$.
Summing over $p \in \Lambda^m$ yields
\begin{equation}
\begin{aligned}
& \sum_{p \in \Lambda^m} \Bigg( \norm{\BQ_J[G]\BU[G][p]\Bf - \BQ_J[\tG]\BU[\tG][p]\Bf}^2 + \sum_{j > -J} \norm{\BQ_j[G]\BU[G][p]\Bf - \BQ_j[\tG]\BU[\tG][p]\Bf}^2 \Bigg) \\
~\leq~ & 3 \sum_{p \in \Lambda^m} \Bigg( \frac{C}{N} \norm{\BU[G][p]\Bf}^2 + \norm{\BU[G][p]\Bf-\BU[\tG][p]\Bf}^2 \Bigg) ~.
\end{aligned}
\end{equation}
That is,
\begin{equation}\label{eq:sumOverP}
\begin{aligned}
& \sum_{p \in \Lambda^m} \norm{\BQ_J[G]\BU[G][p]\Bf - \BQ_J[\tG]\BU[\tG][p]\Bf}^2 + \sum_{p \in \Lambda^{m+1}} \norm{\BU[G][p]\Bf - \BU[\tG][p]\Bf}^2 \\
~\leq~ & \frac{3C}{N} \sum_{p \in \Lambda^m} \norm{\BU[G][p]\Bf}^2 + 3 \sum_{p \in \Lambda^m} \norm{\BU[G][p]\Bf - \BU[\tG][p]\Bf}^2 ~.
\end{aligned}
\end{equation}

To make the following estimation clear,  we denote for $m \geq 1$
\begin{equation*}
\begin{aligned}
a_m ~=~ & \sum_{p \in \Lambda^{m-1}} \norm{\BQ_J[G]\BU[G][p]\Bf - \BQ_J[\tG]\BU[\tG][p]\Bf}^2 ~, \\
b_m ~=~ & \sum_{p \in \Lambda^m} \norm{\BU[G][p]\Bf}^2 ~, \\
d_m ~=~ & \sum_{p \in \Lambda^m} \norm{\BU[G][p]\Bf-\BU[\tG][p]\Bf}^2 ~.
\end{aligned}
\end{equation*}
Also, we denote $b_0 = \norm{f}^2$ and $d_0 = 0$.
Note that $b_m \leq b_0$ for all $m \in \NN \cup \{0\}$. Now (\ref{eq:sumOverP}) can be written as
\begin{equation}
a_m + d_m \leq \frac{3C}{N} b_{m-1} + 3 d_{m-1} ~,~ m \geq 1.
\end{equation}
Summing over $m = 1, \cdots, M$ yields
\begin{equation}
\begin{aligned}
\sum_{m=1}^M a_m ~=~ & \frac{3C}{N} \sum_{m=1}^M b_{m-1} + 3 \sum_{m=1}^M d_{m-1} - \sum_{m=1}^M d_m \\
~\leq~ & \frac{3CM}{N} b_0 + 2 \sum_{m=1}^{M-1} d_m - d_M \\
~\leq~ & \frac{3CM}{N} b_0 + 2 \sum_{m=1}^{M-1} d_m ~.
\end{aligned}
\end{equation}
Note that $d_m \leq 3CN^{-1} b_0 + 3 d_{m-1}$ for $m \geq 1$, and $d_0 = 0$, and hence
\begin{equation}
d_m \leq \Bigg( \frac{1}{2} - \frac{1}{2 \cdot 3^m} \Bigg) \frac{3CM}{N} b_0 ~.
\end{equation}
Therefore,
\begin{equation}
\begin{aligned}
\sum_{m=1}^M a_m ~\leq~ & \frac{3CM}{N} b_0 + 2 \sum_{m=1}^{M-1} \Bigg( \frac{1}{2} - \frac{1}{2 \cdot 3^m} \Bigg) \frac{3CM}{N} b_0 \\
~=~ & \frac{3CM}{N} b_0 + \frac{3CM(M-1)}{N} b_0 - \sum_{m=1}^{M-1} \frac{1}{3^m} \frac{3CM}{N} b_0 \\
~=~ & \frac{3CM^2}{N} b_0 - \frac{3CM}{N} b_0 \left( \frac{1}{2} - \frac{1}{2 \cdot 3^{M-1}} \right) \\
~=~ & \frac{3CM}{N} \left( M - \frac{1}{2} + \frac{1}{2 \cdot 3^{M-1}} \right) b_0 \\
~=~ & \frac{C'}{N} b_0 ~,
\end{aligned}
\end{equation} 
where $C' = 3CM \left( M - 2^{-1} + 2^{-1} 3^{1-M} \right)$. That is, for $\cP_M = \cup_{m=0}^{M-1} \Lambda^m$, the collection of all paths of length smaller than $M$,
\begin{equation}
\sum_{p \in \cP_M} \norm{\BS[G][p]\Bf - \BS[\tG][p]\Bf}^2 \leq \frac{C'}{N} \norm{\Bf}^2 ~.
\end{equation}

On the other hand, summation over paths in $\cP_J \backslash \cP_M$ results in
\begin{equation}
\sum_{p \in \cP_J \backslash \cP_M}  \norm{\BS[G][p]\Bf - \BS[\tG][p]\Bf}^2 \leq 2 \sum_{p \in \cP_J \backslash \cP_M} \Bigg( \norm{\BS[G][p]\Bf}^2 + \norm{\BS[\tG][p]\Bf}^2 \Bigg) \leq \frac{4 C_0}{N} \norm{\Bf}^2 ~.
\end{equation}
Therefore,
\begin{equation}
\norm{\BS[G][\cP_J]\Bf - \BS[\tG][\cP_J]\Bf}^2 \leq \frac{C'+4C_0}{N} \norm{\Bf}^2 ~.
\end{equation}
The generalization to $\BF \in \CC^{N \times D}$  is immediate since $\BF = (\Bf_1, \cdots, \Bf_D)$ and $\norm{\BF}_{\F}^2 = \sum_{d=1}^D \norm{\Bf_d}^2$.

\subsection{Proof of Theorem \ref{thm:weightperturb}}
{This theorem is actually a corollary of Theorem~\ref{thm:forgeneral}. In order to prove it, we only need to show that \eqref{eq:eigenConditions} holds under the assumptions of Theorem \ref{thm:weightperturb}. We define $\BvE := \BL - \tilde{\BL}$ and conclude from \eqref{eq:weak_cond_in_thm} and the definition of the graph Laplacian that  
\begin{equation}
\label{eq:cond_eps_diff}
    \abs{\BvE(n,m)} \leq C_{\sharp} N^{-2}, \textup{ for } n \neq m, \ 1 \leq n,m \leq N  ~~\textup{ and }~~ \BvE(n,n) = - \sum_{m \neq n} \BvE(n,m), \textup{ for } 1 \leq n \leq N ~.
\end{equation}}

{To derive a bound for the perturbation of the eigenvalues, we need to control  \[\norm{\BvE} = \max_{\Bx \neq 0} \frac{\Bx^* \BvE \Bx}{\norm{\Bx}^2} ~.\] 
By applying basic algebraic manipulations as well as the two parts of \eqref{eq:cond_eps_diff}, We obtain that 
\begin{equation*}
    \begin{aligned}
    \Bx^* \BvE \Bx ~=~ & - \sum_{n=1}^N \abs{x_n}^2 \sum_{m \neq n} \BvE(n,m) + \sum_{n=1}^N \sum_{m \neq n} \bar{x}_n \BvE(n,m) x_m \\ 
    ~=~ & - \frac{1}{2} \sum_{n=1}^N \sum_{m \neq n} \abs{x_n - x_m}^2 \BvE(n,m) \\
    ~\leq~ & \frac{1}{2} \max_{\substack{1 \leq n,m \leq N, \\ n \neq m}} \abs{\BvE(n,m)} \cdot 2 \sum_{\substack{1 \leq n,m \leq N, \\ n \neq m}} \abs{x_n}^2 + \abs{x_m}^2 \\
    ~\leq~ & C_{\sharp} N^{-2} (N-1) \norm{\Bx}^2 \\
    ~\leq~ & C_{\sharp} N^{-1} \norm{\Bx}^2 ~.
    \end{aligned}
\end{equation*}
Therefore, $\norm{\BvE} \leq C_{\sharp} N^{-1}$. By Weyl's inequality,
\begin{equation}
    \abs{\l_l - \tilde{\l}_l} \leq \norm{\BvE} \leq C_{\sharp} N^{-1}, \quad l = 0, \cdots, N-1 ~.
\end{equation}}

{Next, we establish bounds for the perturbation of eigenvectors. First note that
\begin{equation*}
    \min_{{l_1} \neq {l_2}} \abs{\l_{l_1} - \tilde{\l}_{l_2}} \geq \min_{{l_1} \neq {l_2}} \left( \abs{\l_{l_1} - \l_{l_2}} - |\l_{l_2} - \tilde{\l}_{l_2}| \right) \geq \delta - C_{\sharp} N^{-1} \geq \delta/2 ~.
\end{equation*}
Let $P_{V_l}$ and $P_{\tilde{V}_l}$ be the orthogonal projections onto the eigenspaces corresponding to $\l_l$ and $\tilde{\l}_l$, respectively. According to the Davis-Kahan Theorem \cite{DavK70, YuWS14}, 
\begin{equation}
    \norm{P_{V_l} - P_{\tilde{V}_l}} \leq \frac{\norm{\BvE}}{\min_{\l_{l_1} \neq \l_{l_2}} \abs{\l_{l_1} - \tilde{\l}_{l_2}}} \leq \frac{2 C_{\sharp}}{N \delta}.
\end{equation}
As a result,
\begin{equation}
    \sin \angle (\Bu_l, \tilde{\Bu}_l) = \norm{\Bu \Bu^* - \tilde{\Bu} \tilde{\Bu}^*} \leq \norm{P_{V_l} - P_{\tilde{V_l}}} \leq 2 C_{\sharp} N^{-1} \delta^{-1} ~.
\end{equation}
We thus note that \eqref{eq:eigenConditions} is satisfied with $C_2 = C_{\sharp}$ and $C_3 = 2 C_{\sharp}/\delta$ and consequently conclude the proof.
}

\section{Numerical results}\label{sec:numerical}
We 
demonstrate the effectiveness of the graph wavelet scattering transform
by using the MNIST dataset for the problem of image classification and the CORA citation network for the problem of community detection. All tests in this section are executed on a PC with Intel i7-6700 CPU, 8GB RAM and GTX1060 6GB GPU. 

\subsection{Image Classification using MNIST}\label{sec:mnist}
The MNIST dataset \cite{LecB98} 
contains $28 \times 28$ gray-scaled images of digits from 0 to 9. 
There are 60,000 training images and 10,000 testing images in total, where the task is to classify the images according to the digits.

In order to test a graph-based method on this dataset, we follow \cite{DefBV16} and construct a graph representing the underlying grid of the images. More specifically, the vertices of this graph correspond to the $28 \times 28$ pixels of each image. For vertices $v_i$ and $v_j$ we let $\dist(v_i,v_j)$ denote the scaled Euclidean distance between the centers of the corresponding pixels so that if $v_i$ and $v_j$ are nearest pixels then $\dist(v_i,v_j)=1$. Edges are drawn between any vertices $v_i$ and $v_j$ satisfying $\dist(v_i,v_j) \leq \sqrt{2}$. That is, each pixel is connected to the nearest pixels in horizontal, vertical and diagonal directions. 
The weight $e^{-\dist(v_i,v_j)^2}$ is assigned to any pair of vertices $v_i$ and $v_j$ connected by an edge. The rest of the pairs have zero weight. 

\begin{figure}[!ht]
\centering
\begin{subfigure}{.19\textwidth}
	\centering
    \includegraphics[width=\linewidth]{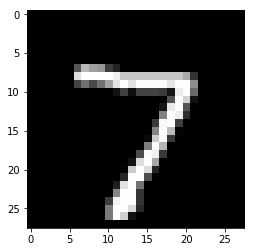}
    \caption{original image}
    \label{fig:origin7}
\end{subfigure}
\begin{subfigure}{.19\textwidth}
	\centering
    \includegraphics[width=\linewidth]{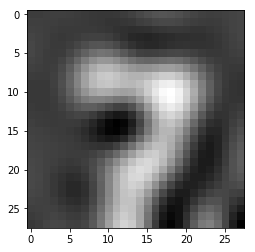}
    \caption{1st layer feature}
    \label{fig:scat71st}
\end{subfigure}
\begin{subfigure}{.19\textwidth}
	\centering
    \includegraphics[width=\linewidth]{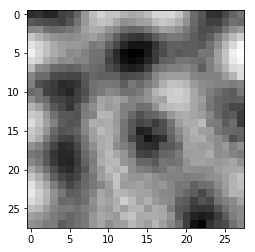}
    \caption{2nd layer feature}
\end{subfigure}
\begin{subfigure}{.19\textwidth}
	\centering
    \includegraphics[width=\linewidth]{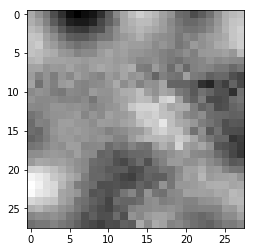}
    \caption{3rd layer feature}
\end{subfigure}
\begin{subfigure}{.19\textwidth}
	\centering
    \includegraphics[width=\linewidth]{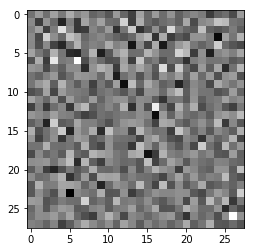}
    \caption{4th layer feature}
    \label{fig:scat74th}
\end{subfigure}
\caption{Examples of features obtained in each layer of the graph scattering network.
The input image $\Bf$ is shown in \ref{fig:origin7}. The scattering transform is applied with $J=3$. The features at the four different layers{, that is, $\BS[\emptyset] \Bf$, $\BS[(-2)] \Bf$, $\BS[(-2,-2)] \Bf$ and $\BS[(-2,-2,-2)] \Bf$,} are demonstrated in Subfigures \ref{fig:scat71st}--\ref{fig:scat74th}.
The pixels of the output images representing the features are arranged in the same manner as in the input image.  
}
\label{fig:mnistFeatures}
\end{figure}

Using this graph, we apply our proposed graph scattering transform with $J=3$, either two or three layers {and the Shannon wavelets}. {The dimension of the output of the scattering transform is $28 \times 28 \times (1+3+9) = 10,192$. It is reduced by PCA to 1,000.}
Figure \ref{fig:mnistFeatures} illustrates {an input image of the digit 7 with the features obtained at the first few layers of the scattering transform with $J=3$ applied to this image. Note that the ``frequency'' of these features} increases with the number of layers.

We use three different classifiers on the features generated via scattering: (1) support vector machine (SVM), (2) softmax layer and (3) fully-connected network (FCN). We use 6,000 images of the training set for validation (simple holdout validation). The accuracies of all three classifiers with and without the scattering transform with 2 and 3 layers are shown in Table \ref{tab:mnist}. Note that the scattering transform is able to generate features that improve the classification results for all classifiers. Moreover, a three-layer network performs better than a two-layer network. 
Three layers almost exhaust the energy of the input signal, so a deeper network is not necessary.
Indeed, we did not notice any improvement when using a fourth-layer.

\begin{table}[!ht]
\centering
\begin{tabular}{|c|c|c|c|}
\hline
~ & SVM & Softmax & FCN \\
\hline
No initial data processing  & 94.16\% & 91.78\% & 98.10\% \\
\hline
Graph scattering transform with $M=2$ layers & 95.68\% & 94.31\% & {99.02}\% \\
\hline
Graph scattering transform with $M=3$ layers & 96.59\% & 94.62\% & {99.09}\%\\
\hline
\end{tabular}
\caption{Classification results on MNIST with and without graph scattering. The first row shows percentage of correct classification by direct application of three common classifiers. The next two lines show classification percentages after preprocessing the data by the graph scattering transform with 2 and 3 layers. }
\label{tab:mnist}
\end{table}

Our best result does not compete with the state-of-the-art result for the MNIST dataset {that obtains 99.75\% accuracy rate \cite{SabFH17}.}
{While the network structure of the method in \cite{SabFH17} is very carefully designed and architected, the graph model used here is only able to encode the information for neighboring pixels. On the other hand, for a convolutional neural network, the convolution at the lowest level collects local information that is not restricted to direct neighbors and is thus able to learn more meaningful local relations.}

{Although the grid graph is not the best way to fully represent the image information, it is still a common benchmark for sanity check of a graph neural network.} Table \ref{sec:wgcn} lists classification results of other graph-based methods. {Results of the first three methods from \cite{EdwX16, HecCQ17, DefBV16} are indicated in parenthesis since they are copied from their original works (codes were not available for the methods of \cite{EdwX16} and \cite{HecCQ17} and the code for Spline filters \cite{DefBV16} did not converge on our computer).
We remark that codes for the methods of \cite{DefBV16} in these experiments and the ones below were obtained from \url{https://github.com/mdeff/cnn\_graph}. }
It is evident that {in terms of accuracy} the scattering transform is comparable with the best graph-based performer.

\begin{table}[!ht]
\centering
\begin{tabular}{|c|c|}
\hline
Method & Accuracy \\
\hline
 Laplacian eigenvalues \cite{EdwX16} & (94.96\%) \\
\hline
 Intuitive convolution \cite{HecCQ17} & (98.55\%) \\
\hline
 Spline filters \cite{DefBV16} & {(97.15\%)} \\
\hline
 Chebyshev filters \cite{DefBV16} & {99.12}\% \\
 \hline
Scattering transform &  {99.09}\% \\
\hline
\end{tabular}
\caption{Percentages of correct classification of different graph-based methods on the MNIST database. 
{Results in parenthesis are copied from their original publications as explained in the main text.}}
\label{tab:mnistcpr}
\end{table}

{To further compare the methods in \cite{DefBV16} and our method, we list the running time for each method on our machine. We note that although the method that uses Chebyshev filters is accurate, it is not computationally efficient. Furthermore, the method that uses spline filters did not converge (DNC) on our computer. On the other hand, the scattering transform achieves a competitive accuracy with high efficiency. }

\begin{table}[!ht]
\centering
\begin{tabular}{|c|c|c|c|}
\hline
~ & Accuracy & Time for scattering & Time for training \\
\hline
Spline filters  & DNC & not needed & 11 s/epoch \\
\hline
Chebyshev filters  & 99.12\% & not needed & 56 s/epoch \\
\hline
Scattering $M=2$ & 99.02\% & 17 s & 2 s/epoch \\
\hline
Scattering $M=3$ & 99.09\% & 36 s & 2 s/epoch \\
\hline
\end{tabular}
\caption{Accuracy and time needed for training MNIST on our machine.  }
\label{tab:mnisttime}
\end{table}

\subsection{Community detection using CORA}\label{sec:SBM}

The CORA dataset \cite{SenNB08} contains 2,708 research papers with 1,433 features describing each paper. There are also 5,429 citation links of the different papers. 
This dataset gives rise to a graph whose vertices correspond to the research papers and edges correspond to citations. We assume an undirected graph, where the weight between two papers is one if at least one of them cite the other, and zero otherwise. 
There are 7 communities of papers and the problem is to detect them. The dataset in \cite{SenNB08} provides labels {(in $\{1,2,\cdots,7\}$)} of 140 vertices for training, 500 vertices for validation, and 1,000 vertices for testing. Due to the small fraction of training samples, the community detection problem in this setting can be considered as semi-supervised learning. The graph scattering transform is applied to the $2,708 \times 1,433$ feature matrix 
with $J=3${,} three layers {and the Shannon wavelets. The dimension of the output of the scattering transform is $1,433 \times (1+3+9) = 18,629$.} The communities are detected by applying FCN to the features obtained by the scattering transform. {We remark that since training with only 140 samples is fast, there is no need to reduce dimension.}

Table \ref{tab:cora} lists the accuracy of the graph scattering transform compared with the state-of-the-art graph-based neural network methods. Note that they are comparable, where the scattering transform demonstrates a slight improvement. All of them outperform the traditional methods listed in \cite[Table 2]{KipW16} (the accuracies of those methods are in the range 59\% -- 75\%).

\begin{table}[!ht]
\centering
\begin{tabular}{|c|c|}
\hline
Method & Accuracy \\
\hline
Chebyshev filters \cite{DefBV16,KipW16} & 79.5\% \\
\hline
Renormalization \cite{KipW16} & 81.5\% \\
\hline
Graph scattering + FCN &  81.9\%  \\
\hline
\end{tabular}
\caption{Percentages of correct labels on CORA for graph scattering and two state-of-the-art methods.}
\label{tab:cora}
\end{table}

\section{Conclusion}

{We constructed a graph convolutional neural network by adapting the scattering transform to graphs. We showed that, with the proper choice of graph wavelets, the graph scattering transform is invariant to permutations and stable to signal and graph manipulations. These invariance and stability properties make the graph scattering transform effective for classification and community detection tasks.}

{Although we exemplified the performance of the graph scattering transform in only two particular instances, where one is a bit artificial, it is a generic tool for feature extraction on graphs. Its potential use is thus not limited to the discriminative tasks illustrated in these two examples. Furthermore, the graph scattering transform does not require training. However, it can be adapted to different datasets and choosing  different kinds of wavelets. In the numerical experiments of this paper we only used the simple Shannon wavelets.}

{In addition to our work, there are other models that try to use the idea of the scattering transform for graphs. For example, the deep Haar scattering \cite{CheCM14}. We believe that our proposed graph scattering network has a more flexible design. Its established  permutation invariance and stability to signal and graph manipulations makes it a robust feature extractor that is natural for graph representation.
The convolutions with the wavelets of \cite{HamVG11} used in our graph transform are somewhat similar to the ones used in trained graph convolutional neural networks such as \cite{DefBV16} and \cite{KipW16}. Our work thus suggest some conceptual understanding of invariance and stability properties for other graph convolutional networks.}

{Despite the advocated properties of the graph scattering transform, it has some limitations. First of all, it is based on the full spectral decomposition of the graph Laplacian and for very large graphs, its computation is demanding.
In order to improve efficiency for the training component, dimension reduction techniques can be used after computing  the graph scattering transform. Second of all, the ``high frequency'' information for the graph Laplacian is not as clear as the high-frequency information in the Euclidean case. Therefore, we do not sufficiently understand the kind of information being processed at deeper layers of the graph scattering transform. At last, the graph scattering transform is a basic generic tool and it may take some time for practitioners to evaluate its potential use. The examples demonstrated here are very simple and the stylized application of classification of images via graph neural networks cannot result in sufficiently competitive results.}

\subsection*{Acknowledgement}
This research was partially supported by NSF {awards DMS-14-18386, DMS-18-21266 and DMS-18-30418}. We thank Radu Balan, Addison Bohannon and Maneesh Singh for helpful references and Loren Anderson, Vahan Huroyan and Tyler Maunu for commenting on an earlier version of this manuscript.

\bibliographystyle{ieeetr}
\bibliography{mainRef}

\end{document}